\documentclass[12pt]{article}

\usepackage{epsfig}
\usepackage{amsmath}
\usepackage{hhline}
\usepackage{amssymb}
\usepackage{times}
\usepackage{cite}
\newlength{\dinwidth}
\newlength{\dinmargin}
\begin{document}  
%%%%%%%%%%%%%%%% Pre-defined commands, you can use for the most obvious notations
\newcommand{\pom}{{I\!\!P}}
\newcommand{\reg}{{I\!\!R}}
\newcommand{\slowpi}{\pi_{\mathit{slow}}}
\newcommand{\fiidiii}{F_2^{D(3)}}
\newcommand{\fiidiiiarg}{\fiidiii\,(\beta,\,Q^2,\,x)}
\newcommand{\n}{1.19\pm 0.06 (stat.) \pm0.07 (syst.)}
\newcommand{\nz}{1.30\pm 0.08 (stat.)^{+0.08}_{-0.14} (syst.)}
\newcommand{\fiidiiiful}{F_2^{D(4)}\,(\beta,\,Q^2,\,x,\,t)}
\newcommand{\fiipom}{\tilde F_2^D}
\newcommand{\ALPHA}{1.10\pm0.03 (stat.) \pm0.04 (syst.)}
\newcommand{\ALPHAZ}{1.15\pm0.04 (stat.)^{+0.04}_{-0.07} (syst.)}
\newcommand{\fiipomarg}{\fiipom\,(\beta,\,Q^2)}
\newcommand{\pomflux}{f_{\pom / p}}
\newcommand{\nxpom}{1.19\pm 0.06 (stat.) \pm0.07 (syst.)}
\newcommand {\gapprox}
   {\raisebox{-0.7ex}{$\stackrel {\textstyle>}{\sim}$}}
\newcommand {\lapprox}
   {\raisebox{-0.7ex}{$\stackrel {\textstyle<}{\sim}$}}
\def\gsim{\,\lower.25ex\hbox{$\scriptstyle\sim$}\kern-1.30ex%
\raise 0.55ex\hbox{$\scriptstyle >$}\,}
\def\lsim{\,\lower.25ex\hbox{$\scriptstyle\sim$}\kern-1.30ex%
\raise 0.55ex\hbox{$\scriptstyle <$}\,}
\newcommand{\pomfluxarg}{f_{\pom / p}\,(x_\pom)}
\newcommand{\dsf}{\mbox{$F_2^{D(3)}$}}
\newcommand{\dsfva}{\mbox{$F_2^{D(3)}(\beta,Q^2,x_{I\!\!P})$}}
\newcommand{\dsfvb}{\mbox{$F_2^{D(3)}(\beta,Q^2,x)$}}
\newcommand{\dsfpom}{$F_2^{I\!\!P}$}
\newcommand{\gap}{\stackrel{>}{\sim}}
\newcommand{\lap}{\stackrel{<}{\sim}}
\newcommand{\fem}{$F_2^{em}$}
\newcommand{\tsnmp}{$\tilde{\sigma}_{NC}(e^{\mp})$}
\newcommand{\tsnm}{$\tilde{\sigma}_{NC}(e^-)$}
\newcommand{\tsnp}{$\tilde{\sigma}_{NC}(e^+)$}
\newcommand{\st}{$\star$}
\newcommand{\sst}{$\star \star$}
\newcommand{\ssst}{$\star \star \star$}
\newcommand{\sssst}{$\star \star \star \star$}
\newcommand{\tw}{\theta_W}
\newcommand{\sw}{\sin{\theta_W}}
\newcommand{\cw}{\cos{\theta_W}}
\newcommand{\sww}{\sin^2{\theta_W}}
\newcommand{\cww}{\cos^2{\theta_W}}
\newcommand{\trm}{m_{\perp}}
\newcommand{\trp}{p_{\perp}}
\newcommand{\trmm}{m_{\perp}^2}
\newcommand{\trpp}{p_{\perp}^2}
\newcommand{\alp}{\alpha_s}

\newcommand{\alps}{\alpha_s}
\newcommand{\sqrts}{$\sqrt{s}$}
\newcommand{\LO}{$O(\alpha_s^0)$}
\newcommand{\Oa}{$O(\alpha_s)$}
\newcommand{\Oaa}{$O(\alpha_s^2)$}
\newcommand{\PT}{p_{\perp}}
\newcommand{\JPSI}{J/\psi}
\newcommand{\sh}{\hat{s}}
\newcommand{\uh}{\hat{u}}
\newcommand{\MP}{m_{J/\psi}}
\newcommand{\PO}{I\!\!P}
\newcommand{\xbj}{x}
\newcommand{\xpom}{x_{\PO}}
\newcommand{\ttbs}{\char'134}
\newcommand{\xpomlo}{3\times10^{-4}}  
\newcommand{\xpomup}{0.05}  
\newcommand{\dgr}{^\circ}
\newcommand{\pbarnt}{\,\mbox{{\rm pb$^{-1}$}}}
\newcommand{\gev}{\,\mbox{G\eV}}
\newcommand{\WBoson}{\mbox{$W$}}
\newcommand{\fbarn}{\,\mbox{{\rm fb}}}
\newcommand{\fbarnt}{\,\mbox{{\rm fb$^{-1}$}}}
\newcommand{\dsdx}[1]{$d\sigma\!/\!d #1\,$}
\newcommand{\eV}{\mbox{e\hspace{-0.08em}V}}
%
% Some useful tex commands
%
\newcommand{\qsq}{\ensuremath{Q^2} }
\newcommand{\gevsq}{\ensuremath{\mathrm{G\eV}^2} }
\newcommand{\et}{\ensuremath{E_t^*} }
\newcommand{\rap}{\ensuremath{\eta^*} }
\newcommand{\gp}{\ensuremath{\gamma^*}p }
\newcommand{\dsiget}{\ensuremath{{\rm d}\sigma_{ep}/{\rm d}E_t^*} }
\newcommand{\dsigrap}{\ensuremath{{\rm d}\sigma_{ep}/{\rm d}\eta^*} }

% BtoMuX stuff
\newcommand{\ptjetone}{\ensuremath{P_{T}^{\text{jet1}}}}
\newcommand{\ptjettwo}{\ensuremath{P_{T}^{\text{jet2}}}}
\newcommand{\ptmuon}{\ensuremath{P_{T}^{\mu}}}
\newcommand{\ptmuonjet}{\ensuremath{P_{T}^{\mu\text{jet}}}}
\newcommand{\ptotherjet}{\ensuremath{P_{T}^{\text{other jet}}}}
\newcommand{\etajet}{\ensuremath{\eta^{\text{jet}}}}
\newcommand{\etamuon}{\ensuremath{\eta^{\mu}}}
\newcommand{\phimuon}{\ensuremath{\phi^{\mu}}}
\newcommand{\ptrel}{\ensuremath{P_{T}^{rel}}}
\newcommand{\muonjet}{\ensuremath{\mu\text{-jet}}}
\newcommand{\xgamma}{\ensuremath{x_{\gamma}^{\text{obs}}}}
\newcommand{\dphijets}{\ensuremath{\Delta\phi^{\text{jj}}}}
\newcommand{\impactpar}{\ensuremath{\delta}}

%%% Dstar stuff
\newcommand{\dstar}{\ensuremath{D^*}}
\newcommand{\dstarp}{\ensuremath{D^{*+}}}
\newcommand{\dstarm}{\ensuremath{D^{*-}}}
\newcommand{\dstarpm}{\ensuremath{D^{*\pm}}}
\newcommand{\zDs}{\ensuremath{z(\dstar )}}
\newcommand{\Wgp}{\ensuremath{W_{\gamma p}}}
\newcommand{\ptds}{\ensuremath{p_t(\dstar )}}
\newcommand{\etads}{\ensuremath{\eta(\dstar )}}
\newcommand{\ptj}{\ensuremath{p_t(\mbox{jet})}}
\newcommand{\ptjn}[1]{\ensuremath{p_t(\mbox{jet$_{#1}$})}}
\newcommand{\etaj}{\ensuremath{\eta(\mbox{jet})}}
\newcommand{\detadsj}{\ensuremath{\eta(\dstar )\, \mbox{-}\, \etaj}}

% Journal macro
\def\Journal#1#2#3#4{{#1} {\bf #2} (#3) #4}
\def\NCA{\em Nuovo Cimento}
\def\NIM{\em Nucl. Instrum. Methods}
\def\NIMA{{\em Nucl. Instrum. Methods} {\bf A}}
\def\NPB{{\em Nucl. Phys.}   {\bf B}}
\def\PLB{{\em Phys. Lett.}   {\bf B}}
\def\PRL{\em Phys. Rev. Lett.}
\def\PRD{{\em Phys. Rev.}    {\bf D}}
\def\ZPC{{\em Z. Phys.}      {\bf C}}
\def\EJC{{\em Eur. Phys. J.} {\bf C}}
\def\CPC{\em Comp. Phys. Commun.}

\begin{titlepage}

\noindent
\begin{flushleft}
{\tt DESY 12-059    \hfill    ISSN 0418-9833} \\
{\tt April 2012}                  \\
\end{flushleft}

\vspace{2cm}
\begin{center}
\begin{Large}

{\bf Measurement of Beauty and Charm Photoproduction using Semi-muonic Decays in Dijet Events at HERA }

\vspace{1cm}%2cm

H1 Collaboration

\end{Large}
\end{center}

\vspace{1cm} %2cm

\begin{abstract}
\noindent Measurements of cross sections for beauty and charm events with dijets and a muon in the photoproduction regime at HERA are presented. The data were collected with the H1 detector and correspond to an integrated luminosity of $179$ pb$^{-1}$. Events with dijets of transverse momentum \ptjetone $> 7$ G\eV \ and \ptjettwo $> 6$ G\eV \ in the pseudorapidity range $-1.5<\etajet < 2.5$ in the laboratory frame are selected in the kinematic region of photon virtuality \qsq $< 2.5$ \gevsq and inelasticity $0.2 < y < 0.8$. One of the two selected jets must be associated to a muon with \ptmuon $> 2.5$ G\eV\  in the pseudorapidity range $-1.3 <$ \etamuon $< 1.5$.  The fractions of beauty and charm events are determined using  the impact parameters of the muon tracks with respect to the primary vertex and their transverse momentum relative to the axis of the associated jet. 
Both variables are reconstructed using the H1 vertex detector. The measurements are in agreement with QCD predictions at leading and next-to-leading order. 
\end{abstract}

\vspace{0.5cm}%1.5cm

\begin{center}
Submitted to \EJC \;\; 
\end{center}

\end{titlepage}

%          THE PAPER DRAFTS HAVE NO AUTHORLIST
%
%          FOR PAPER ISSUED FOR THE FINAL READING 
%          COPY THE AUTHOR AND INSTITUTE LISTS 
%          INTO YOUR AREA
%
% from /h1/iww/ipublications/h1auts.tex 
%          AND UNCOMMENT THE NEXT THREE LINES 
%
\begin{flushleft}
  % \input{h1auts}
%-- H1AUTS Author list by names 
%-- Status: Tue Apr  3 13:47:10 CEST 2012  Number of authors = 197 

F.D.~Aaron$^{5,48}$,           %BUCH-LEFT      01/12           Aaron               
C.~Alexa$^{5}$,                %BUCH-PD        06/06           Alexa               
V.~Andreev$^{25}$,             %LPI -PD        8/88            Andreev             
S.~Backovic$^{30}$,            %PODG-LEFT      08/11           Backovic            
A.~Baghdasaryan$^{38}$,        %YERE-PD        09/03           Baghdasaryana       
S.~Baghdasaryan$^{38}$,        %YERE-ST        02/10           Baghdasaryans       
E.~Barrelet$^{29}$,            %PARI-HON       01/10           Barrelet            
W.~Bartel$^{11}$,              %DESY-HON       02/03           Bartel              
K.~Begzsuren$^{35}$,           %ULBA-PD        04/06           Begzsuren           
A.~Belousov$^{25}$
% ...

%\newpage

%-- H1 Institutes 
\bigskip{\it
 $ ^{1}$ I. Physikalisches Institut der RWTH, Aachen, Germany \\
% ...
$ ^{5}$ National Institute for Physics and Nuclear Engineering (NIPNE) ,
          Bucharest, Romania$^{ m}$ \\
% ...
$ ^{11}$ DESY, Hamburg, Germany \\
% ...
$ ^{25}$ Lebedev Physical Institute, Moscow, Russia \\
% ...
$ ^{29}$ LPNHE, Universit\'e Pierre et Marie Curie Paris 6,
          Universit\'e Denis Diderot Paris 7, CNRS/IN2P3, Paris, France \\
 $ ^{30}$ Faculty of Science, University of Montenegro,
          Podgorica, Montenegro$^{ n}$ \\
% ...
$ ^{35}$ Institute of Physics and Technology of the Mongolian
          Academy of Sciences, Ulaanbaatar, Mongolia \\
% ...
$ ^{38}$ Yerevan Physics Institute, Yerevan, Armenia \\
% ...
$ ^{48}$ Also at Faculty of Physics, University of Bucharest,
          Bucharest, Romania \\
% ...
$ ^m$ Supported by the Romanian National Authority for Scientific Research
      under the contract PN 09370101 \\
 $ ^n$ Partially Supported by Ministry of Science of Montenegro,
      no. 05-1/3-3352 \\
}
\end{flushleft}
%
% Please not that the author list may need re-formatting.

%\newpage

%\tableofcontents

\newpage

%%%%%%%%%%%%%%%%%%%%%%%%%%%%%%%%%%%%%%%%%%%%%%%%%%%%%%%%%%
\section{Introduction}
%%%%%%%%%%%%%%%%%%%%%%%%%%%%%%%%%%%%%%%%%%%%%%%%%%%%%%%%%%
The photoproduction of heavy quarks at the HERA electron-proton collider is of
particular interest for testing calculations in the framework of perturbative
quantum chromodynamics (pQCD). The masses $m_b$ and $m_c$ of the beauty and
charm quarks, as well as the transverse momentum of a jet, provide a hard scale,
which is necessary for the calculation of pQCD predictions \cite{mcatnlo}. The
dominant process for beauty and charm production at HERA is boson-gluon fusion
(BGF): $\gamma g \rightarrow Q\bar{Q}X$, with $Q=b, c$. In the kinematic regime
of photoproduction, where the photon virtuality \qsq is small, two classes of processes contribute to BGF. In direct photon processes, the photon emitted from the electron\footnote{In this paper the term 'electron' is used generically to refer to both electrons and positrons.} enters the hard BGF process directly. In resolved photon processes, the photon fluctuates into a hadronic state before the hard interaction and acts as a source of partons, one of which takes part in the hard interaction. At leading order (LO) pQCD resolved photon processes are expected to contribute significantly in the photoproduction region.
\par
Heavy quark photoproduction has been investigated at HERA using different analysis techniques in different regions of phase space. The lifetime or mass of heavy flavoured hadrons \cite{ZeusLifetime, H1Lifetime2010, H1Lifetime2011, H1Hera1desy05004, ZEUSHera1desy03212,ZEUSdesy08210}, semileptonic decays \cite{ZeusSemileptonic, H1Hera1desy05004, ZEUSHera1desy03212} or the full reconstruction of a $D$ or $D^*$ meson \cite{H1D, ZeusD} are exploited to perform the measurements. In general, the measured cross sections agree reasonably well with the theoretical predictions. In the H1 measurement performed with HERA I data \cite{H1Hera1desy05004}, where beauty photoproduction was investigated using two jets and a muon in the final state, the data were found to be described reasonably well by the pQCD calculations at next-to-leading order (NLO), except in the region of low transverse momentum of the muon $2.5<\ptmuon\ <3.3$ \gev \ and of low leading jet transverse momentum $7.0< \ptjetone <10.0$ \gev, where the NLO prediction was lower than the data. Comparable measurements were performed by the ZEUS collaboration \cite{ZEUSHera1desy03212,ZEUSdesy08210}, covering a similar phase space. Here, a good description by the NLO QCD predictions is found, also at low transverse momenta of the leading jet and the muon. 
\par
This paper reports on beauty and charm measurements of cross sections for
photoproduction of events with two jets and a muon, where the muon is associated
with one of the jets. The analysis exploits the lifetime and the mass of heavy
flavoured hadrons as in the former HERA I beauty production analysis
\cite{H1Hera1desy05004}. The measurement is additionally performed for charm
photoproduction. Increased statistics due to increased luminosity and extended
phase space, a better understanding and description of the H1 vertex detector
\cite{CSTTuning}, which is crucial for this ana\-lysis, as well as reduced
systematic uncertainties compared to the previous H1 ana\-lysis make a more
detailed test of pQCD predictions possible. Going beyond the earlier HERA I
analysis, the cross sections as a function of the azimuthal angular difference \dphijets\  between the two leading jets, which are sensitive to higher order corrections, are also measured. Furthermore, cross sections are measured in two different regions of phase space, which are either enriched by resolved or direct photon processes.

%%%%%%%%%%%%%%%%%%%%%%%%%%%%%%%%%%%%%%%%%%%%%%%%%%%%%%%%%%
\section{QCD Calculations}\label{MonteCarloSimulation}
%%%%%%%%%%%%%%%%%%%%%%%%%%%%%%%%%%%%%%%%%%%%%%%%%%%%%%%%%%
The data presented here are compared with LO calculations supplemented by parton showers as well as with NLO calculations. The calculations are performed using either collinear factorisation, which is based on the DGLAP evolution equations \cite{DGLAP}, or the $k_t$-factorisation approach, which employs the CCFM \cite{CCFM} evolution equations. In the collinear approach transverse momenta obtained through initial state QCD evolution are neglected and all the transverse momenta are generated in the hard scattering process, i.e. the partons entering the hard interaction are collinear with the proton. Effects from non-zero transverse momenta of the gluons enter only at NLO. In the $k_t$-factorisation ansatz the transverse momenta of gluons $k_t$ entering the hard interaction are already included at leading order both in the off-shell matrix element and the $k_t$-dependent unintegrated gluon density \cite{uPDFs}. Therefore, corrections appearing only at higher order in collinear factorisation are partially included at LO in the $k_t$-factorisation approach.
\par
For beauty and charm photoproduction two classes of processes occur, the direct photon processes and the resolved photon processes.
The distinction between these two classes depends on the factorisation scheme and the order in which the calculation is performed. 
\par
The production of heavy quarks is calculated either in the massive scheme, where heavy quarks are produced only perturbatively via boson gluon fusion, or in the massless scheme, where heavy quarks are treated as massless partons. These two schemes are expected to be appropriate in different regions of phase space \cite{schemes}: the massive scheme is expected to be reliable when the transverse momentum $P_T$ of the heavy quarks is of similar size compared to the heavy quark mass $m_Q$, whereas the massless scheme is expected to be valid for $P_T\gg m_Q$.
\par
An overview of the parameters used in the Monte Carlo (MC) generators is given in table \ref{table_qcdsettings}. The following MC generators are used:

\begin{description}
\item{\textsc{Pythia}: The MC program \textsc{Pythia} 6.4 \cite{Pythia} is based on LO QCD matrix elements with leading-log parton showers in the collinear factorisation approach. \textsc{Pythia} includes both direct photon gluon fusion and resolved photon processes. In the resolved photon processes either a beauty or a charm quark or a gluon from the photon enters the hard scattering. In the inclusive mode of \textsc{Pythia} used here beauty and charm quarks are treated as massless partons in all steps of the calculation in both types of processes. Three samples are generated containing photoproduction events for the processes $ep\rightarrow b\mu X$, $ep\rightarrow c\mu X$ and $ep\rightarrow qjX$ where $q$ is a light quark of flavour $u$, $d$ or $s$ and $j$ denotes a jet. The latter sample is generated without specifically requiring a muon in order to use it for studying the background arising from muon candidates which originate from sources other than beauty or charm processes.
The hadronisation process is simulated according to the Lund string model  \cite{LundFrag}. For the systematic uncertainty arising from the fragmentation model, additional samples are generated using the Peterson fragmentation function \cite{PeteFrag} for heavy quarks.
}
%\end{description}

%\begin{description}
\item{\textsc{Cascade}: The MC program \textsc{Cascade} 2.0 \cite{Cascade} is used for simulating events based on LO QCD calculations in the $k_t$-factorisation approach. Only the direct boson gluon fusion process is implemented using off-shell matrix elements. Higher order QCD corrections are simulated with initial state parton showers applying the CCFM evolution \cite{CCFM}. Here, two samples containing the processes $ep\rightarrow b\mu X$ and $ep\rightarrow c\mu X$ are generated. The unintegrated PDFs of the proton from set A0 \cite{CCFMSetA0} are used. The hadronisation of partons is performed with the Lund string model as implemented in \textsc{Pythia}. 
}
%\end{description}

%\begin{description}
\item{\textsc{Herwig}: The MC program \textsc{Herwig} 6.510 \cite{herwig} is used to simulate events based on collinear factorisation and massless LO QCD calculations. \textsc{Herwig} includes both direct photon gluon fusion and resolved photon processes. The hadronisation of partons is performed using the cluster fragmentation~\cite{cluster}. 
}
%\end{description}
%\begin{description}
\item{MC@NLO: The MC@NLO program for HERA \cite{mcatnlo} combines a collinear NLO calculation of heavy flavour production in the massive approach \cite{fmnr} with parton showers and hadronisation as described in \cite{mcatnlohad}. The direct and resolved parts of the cross section are calculated separately. MC@NLO uses parton showers applying the DGLAP evolution to simulate higher order contributions and the cluster fragmentation~\cite{cluster} as implemented in \textsc{Herwig} \cite{herwig}. The quark masses are set to $m_c=1.5$ G\eV \ and $m_b=4.75$ G\eV. The central value of the renormalisation scale $\mu_{R}$ is set to $\sqrt{m_Q^2+P_T^2}$, while the factorisation scale $\mu_F$ is $2\mu_R$. As an estimate of the theoretical uncertainties on the NLO QCD predictions the scales $\mu_R$ and $\mu_F$ are varied separately by factors of $0.5$ and $2$, while $m_b$ is changed by \mbox{$\pm0.25$ G\eV} and $m_c$ is changed by $\pm0.2$ G\eV. The resulting variations are added in quadrature to obtain the resulting theoretical uncertainty. 
}
\end{description}

\renewcommand{\arraystretch}{1.18} % this increase the table in vertical direction, very useful!

\begin{table}

\centering
\begin{tabular}{|c|c|c|c|c|}

\hline
               &\textsc{Pythia} & \textsc{Cascade} & \textsc{Herwig} & MC@NLO \\ \hline \hline
 Version & 6.4 & 2.0 & 6.510 & \textsc{Herwig} 6.510     \\  
 Evolution scheme & DGLAP & CCFM &  DGLAP  &  DGLAP    \\ \hline                          
 $m_b$ [G\eV] & $4.75$ & $4.75$ & $4.75$ & $4.75$  \\
 $m_c$ [G\eV] & $1.5$   & $1.5$ & $1.5$  & $1.5$   \\ \hline
 Proton PDF & CTEQ6L1\cite{CTEQ6}   & A0\cite{CCFMSetA0} & HERAPDF 1.0\cite{HERAPDF} & HERAPDF 1.0   \\
 Photon PDF & SaS2D\cite{Sas2d}   &  -  & GRV-G LO\cite{GRVGLO} & GRV-G LO   \\ \hline
& & & & \\[-0.5cm]
Renorm. scale $\mu_R$ & $\sqrt{P_T^2+\frac{1}{2}(m_1^2+m_2^2)}$  & $\sqrt{\hat{s}+Q_{\bot}^2}$  & $\sqrt{m_Q^2+P_T^2}$  & $\sqrt{m_Q^2+P_T^2}$ \\
Factor. scale $\mu_F$ & $\mu_R$ & $\mu_R$ &$\mu_R$ & $2\mu_R$\\ \hline
Fragmentation & Lund  & Lund & cluster & cluster \\    \hline
                           & $a=0.437$  &  $a=0.437$ & - & - \\
                            &  $b=0.850$ & $ b=0.850$ & - & -\\
                           & Peterson & - & - & - \\
                           & $\epsilon_b=0.006$ & - & - & -\\ 
                          & $\epsilon_c=0.06$ & - & - & -\\ \hline
 \end{tabular}
\caption{Parameters used in the QCD calculations \textsc{Pythia}, \textsc{Cascade}, \textsc{Herwig}, and MC@NLO. Here, $P_T$ stands for the transverse momentum, $m_1, \ m_2$ are the masses of the two outgoing partons from the hard process, and $m_Q$ denotes the heavy quark mass. The variable $\hat{s}$ represents the invariant mass of the $Q\bar{Q}$ system and $Q_{\bot}$ stands for its transverse momentum.}\label{table_qcdsettings}
\end{table}

\textsc{Pythia} and \textsc{Cascade} are used to simulate detector effects in order to determine the acceptance and the efficiency and to estimate the systematic uncertainties associated with the measurement. The generated events are passed through a detailed simulation of the detector response based on the GEANT simulation program \cite{Brun:1987ma} and are processed using the same reconstruction and analysis chain as is used for the data.  

%%%%%%%%%%%%%%%%%%%%%%%%%%%%%%%%%%%%%%%%%%%%%%%%%%%%%%%%%%
\section{H1 Detector}
%%%%%%%%%%%%%%%%%%%%%%%%%%%%%%%%%%%%%%%%%%%%%%%%%%%%%%%%%%
Only a short description of the H1 detector is given here including the most relevant detector components for this analysis. A more complete description may be found elsewhere \cite{H1detector, Appuhn:1996na}. A right-handed coordinate system is employed at H1, with its origin at the nominal interaction vertex, its $z$-axis pointing in the proton beam direction and its $x(y)$ axis pointing in the horizontal (vertical) direction. Polar ($\theta$) and azimuthal ($\phi$) angles are measured with respect to this reference system. The pseudorapidity $\eta$ is related to the polar angle $\theta$ by $\eta=-\ln\tan(\theta/2)$.
\par
Charged particles are measured in the central tracking detector (CTD) with a transverse momentum resolution of $\sigma(P_T)/P_T\approx0.5\%P_T/\text{GeV} \oplus 1.5\%$ \cite{CTDResolution}. This device consists of two cylindrical drift chambers (CJC) interspersed with a drift chamber designed to improve the $z$-coordinate reconstruction. A multiwire proportional chamber mainly used for triggering is located in front of the inner CJC. The CTD is operated in a uniform solenoidal $1.16 \ \text{T}$ magnetic field, enabling the momentum measurement of charged particles over the polar angular range $20^{\circ}<\theta<160^{\circ}$. The efficiency for finding tracks in the CTD is greater than $99\%$.
\par
The CTD tracks are linked to hits in the vertex detector, the central silicon tracker (CST) \cite{CST}, to provide precise spatial track reconstruction. The CST consists of two layers of double-sided silicon strip detectors surrounding the beam pipe, covering an angular range of $30^{\circ}<\theta<150^{\circ}$ for tracks passing through both layers. The information of the $z$-coordinate of the CST hits is not used in the analysis presented in this paper. For CTD tracks with CST hits in both layers the transverse distance of closest approach (DCA) to the nominal vertex in $x-y$, averaged over the azimuthal angle, is measured to have a resolution of $43\ \mu$m $\oplus$ $51\ \mu$m$/(P_T/\text{GeV})$, where the first term represents the intrinsic resolution (including alignment uncertainty) and the second term is the contribution from multiple scattering in the beam pipe and the CST. The efficiency for linking hits in both layers of the CST to a CTD track is around $84\%$. 
\par
The track detectors are surrounded in the forward and central directions ($4^{\circ}<\theta<154^{\circ}$) by a finely grained liquid argon calorimeter (LAr) and in the backward region ($153^{\circ}<\theta<178^{\circ}$) by a lead-scintillating fibre calorimeter (SpaCal) both with electromagnetic and hadronic sections. These calorimeters provide energy and angular reconstruction for final state particles from the hadronic system. In the LAr electromagnetic shower energies are measured with a precision of $\sigma(E)/E=11\%/\sqrt{E/\text{GeV}}\oplus 1\%$ and hadronic energies with $\sigma(E)/E=50\%/\sqrt{E/\text{GeV}} \oplus 1\%$, as determined in test beam measurements. The energy resolution for electromagnetic showers in the SpaCal is $\sigma(E)/E=7.1\%/\sqrt{E/\text{GeV}}\oplus1\%$, as determined in test beam measurements \cite{SpaCalTestBeam}.
\par
The calorimeters are surrounded by the muon system. The central muon detector (CMD) is integrated in the iron return yoke of the superconducting coil and consists of $64$ modules, which are grouped in the forward endcap ($5^{\circ}\leq\theta\leq35^{\circ}$), the forward and backward barrel ($35^{\circ}\leq\theta\leq130^{\circ}$) and the backward endcap ($130^{\circ}\leq\theta\leq175^{\circ}$). Muon candidates are identified by requiring a geometrical matching of a CMD track segment with a CTD track.
\par
The luminosity determination is based on the measurement of the Bethe-Heitler process $ep\rightarrow ep\gamma$, where the photon is detected in a calorimeter located downstream of the interaction point in the electron beam direction at $z=-103\ \text{m}$.

%%%%%%%%%%%%%%%%%%%%%%%%%%%%%%%%%%%%%%%%%%%%%%%%%%%%%%%%%%
\section{Experimental Method}
%%%%%%%%%%%%%%%%%%%%%%%%%%%%%%%%%%%%%%%%%%%%%%%%%%%%%%%%%%
The data were collected with the H1 detector at the HERA collider during the years $2006$ and $2007$ and correspond to an integrated luminosity of $\mathcal{L}=179 \ \text{pb}^{-1}$. 
The beam energies were $E_e=27.6 \ \text{G\eV}$ and $E_p=920 \ \text{G\eV}$ for electrons and protons, respectively, resulting in a centre-of-mass energy of $\sqrt{s}\approx 320 \ \text{G\eV}$.  The  trigger requires a track segment in the muon system and track activity in the central jet chamber. A detailed account of the present analysis can be found in \cite{MiraKraemerThesis}. A summary of the kinematic range and the definition of the measurement is given in table \ref{CutSummaryTable}. 

\subsection{Photoproduction Event Selection}
Events in the photoproduction regime are selected by requiring that no isolated high energy electromagnetic cluster, consistent with a signal from a scattered electron, is detected in the LAr and SpaCal calorimeters. This limits the photon virtuality to values of \qsq$< 2.5 \ \gevsq$. 
The inelasticity $y$ is reconstructed using the relation $y=\sum_h(E-P_z)/2E_e$ \cite{JacquetBlondel}. Here, the sum includes all particles of the hadronic final state (HFS), while $E$ denotes their energies and $P_z$ stands for the $z$-components of their momenta. The HFS particles are reconstructed using a combination of tracks and calorimeter deposits in an energy flow algorithm that avoids double counting \cite{HFS}. The inelasticity in this analysis is restricted to $0.2 < y < 0.8$. 

\subsection{Muon Reconstruction and Selection}\label{subsec:MuonRec}
Muon candidates are identified as track segments in the barrel and endcap parts of the instrumented iron. The iron track segments must be well matched to a track reconstructed in the CTD. At least two CST hits in the $r-\phi$ plane have to be associated with the muon track. The combined CTD-CST track in $r-\phi$ is required to have a fit probability of at least $10\%$. The muon momentum is reconstructed using the CTD-CST track information. The CST hit requirements for the muon track restrict the allowed range of $ep$ interaction vertices along the $z$-axis to $|z_{vtx}|\leqslant 20 \ \text{cm}$. Events are selected with at least one muon candidate reconstructed in the instrumented iron having a pseudorapidity within $-1.3<\etamuon<1.5$ and a transverse momentum of $\ptmuon>2.5 \ \text{G\eV}$. If more than one muon candidate is found, the one with the highest transverse momentum is selected and other candidates are ignored. In $5.4\%$ of the events after the full selection more than one muon is found. 

\subsection{Jet Reconstruction and Selection}
Jets are reconstructed using the inclusive longitudinally invariant $k_T$ algorithm in the massless $P_T$ recombination scheme and with the distance parameter $R_0=1$ in the $\eta-\phi$ plane \cite{JetAlgo}. The algorithm is applied in the laboratory frame using all reconstructed HFS particles including the muon candidate. A jet is defined as a \muonjet\  if the selected muon candidate lies within a cone of radius 1 about the jet axis in the $\eta-\phi$ plane. The efficiency for this matching amounts to about $90\%$ and is consistent for data and MCs.  
The jet with the highest $P_T$ is referred to as jet1, while the second highest is called jet2. Events with at least two jets are selected, where the leading two jets are required to be in the angular range $-1.5<\etajet<2.5$ and to have a transverse momentum of $\ptjetone>7\ \text{G\eV}$ and $\ptjettwo>6\ \text{G\eV}$. One of the two selected jets must be classified as a \muonjet. The Monte Carlo simulation is used to define hadron level jets, which consist of stable particles including neutrinos, but excluding the scattered electron, before they are passed through the simulation of the detector response.

\subsection{Separation of Direct and Resolved Processes}
The fraction of the photon energy entering the hard interaction is estimated using the observable $\xgamma$:
\begin{equation}\nonumber
\xgamma=\frac{\sum_{jet1}(E-P_z)+\sum_{jet2}(E-P_z)}{\sum_{HFS}(E-P_z)},
\end{equation}
where the sums in the numerator run over the particles associated with the two jets and the one in the denominator over all detected hadronic final state particles. For direct processes \xgamma\  approaches unity, as the hadronic final state consists only of the two hard jets selected in the present analysis and the proton remnant in the forward region only has a minor contribution to $\sum_{HFS}(E-P_z)$. In resolved processes \xgamma\  can have smaller values. 

\begin{table}
\centering
\begin{tabular}{|c|c|}
\hline 
& Photoproduction of $b(c)\rightarrow \mu jj X$  \\
\hline
\hline 
Kinematic range &  \qsq$< 2.5 \ \gevsq$  \\
 & $0.2 < y < 0.8$ \\
\hline
Event selection & \ptmuon $>2.5 \ \text{G\eV}$ \\
& $-1.3<\etamuon<1.5$\\
& \ptjetone $> 7 \ \text{G\eV}$ \\
& \ptjettwo $>6 \ \text{G\eV}$\\
& $-1.5<\etajet<2.5$ \\
\hline
Event sample & $N_{events}=6807$ \\
& $\mathcal{L}=179 \ \text{pb}^{-1}$ \\
\hline
\end{tabular}
\caption{Definition of the kinematic range of the measurement and event yield for the data sample collected in the years $2006$ and $2007$. The variables are measured in the laboratory frame.}\label{CutSummaryTable}
\end{table}

\subsection{Flavour Separation}
The flavour of an event is defined as the hadron flavour of the $\muonjet$. The measured cross sections are proportional to the rate of events with a muon and a dijet system rather than the rate of muons or jets. 
\par
The separation of $b$, $c$ and light quark ($uds$) events is only briefly described here. The procedure closely follows that described in \cite{H1Hera1desy05004}. The separation is performed using the properties of the muon track associated to the \muonjet. The impact parameter $\delta$ of a muon track is the transverse distance of closest approach (DCA) of the muon track to the beam spot point, which is the position of the beam interaction region in $x$ and $y$. The beam spot is derived from tracks with CST hits averaging over many events and is updated regularly to account for drifts during beam storage. Muon tracks with $\delta > 0.1 \ \text{cm}$ are rejected to suppress contributions from the decays of long-lived strange particles. If the angle $\alpha$ between the azimuthal angle of the \muonjet \ and the line joining the primary vertex to the point of  muon closest approach is less than $90^{\circ}$, the impact parameter is defined as positive. It is defined as negative otherwise. The transverse momentum \ptrel \ of the muon relative to the \muonjet \ axis is also sensitive to the quark content of the event sample and used together with the impact parameter for the flavour separation.
\par
The fractions of events with beauty, charm and light quarks, $f_{b}$, $f_{c}$ and $f_{l}$, are obtained by a binned likelihood fit \cite{BarlowFit} in the $\impactpar - \ptrel$ plane. Following \cite{BarlowFit}, a likelihood ratio is calculated based on Poisson statistics. The fit is performed separately for each individual measurement bin, while the total cross sections are determined using the fractions obtained from a fit to the complete event sample. The $uds$ (light), $c$ and $b$ \textsc{Pythia} Monte Carlo simulation samples are used as templates. Only the statistical errors of the data and the Monte Carlo simulations are considered in the fit. As a cross check, all fits are also performed using one-dimensional distributions of \ptrel\ and \impactpar\ separately. These two one-dimensional fits give a compatible $b$ fraction in all measurement bins within the statistical error. The one-dimensional \ptrel\ fit does not allow a determination of the charm fraction.
\par
\begin{figure}
\setlength{\unitlength}{1cm}
\centering
\begin{picture}(15.,8.)(0.,0.)
\put(-1.5,0.){\epsfig{file=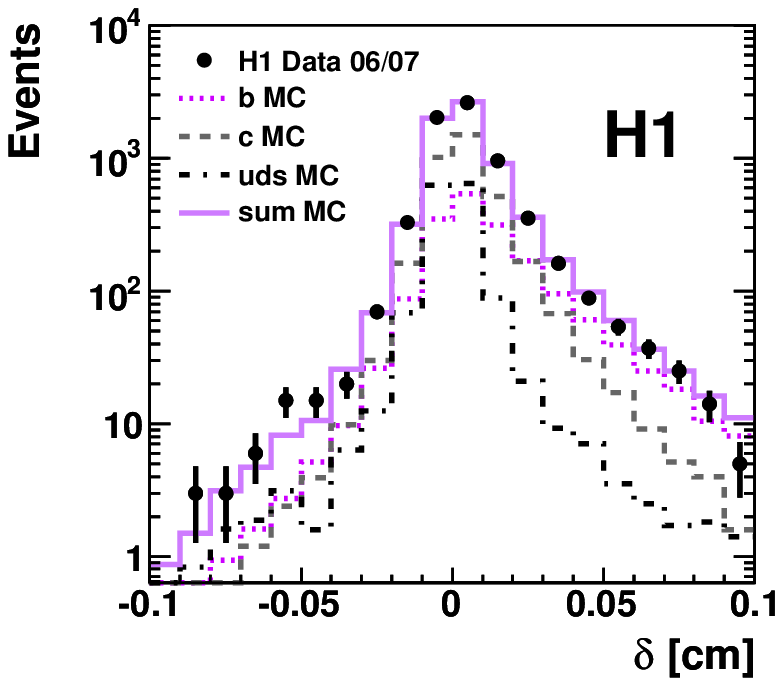,width = 9cm}}
\put(7.7,0.){\epsfig{file=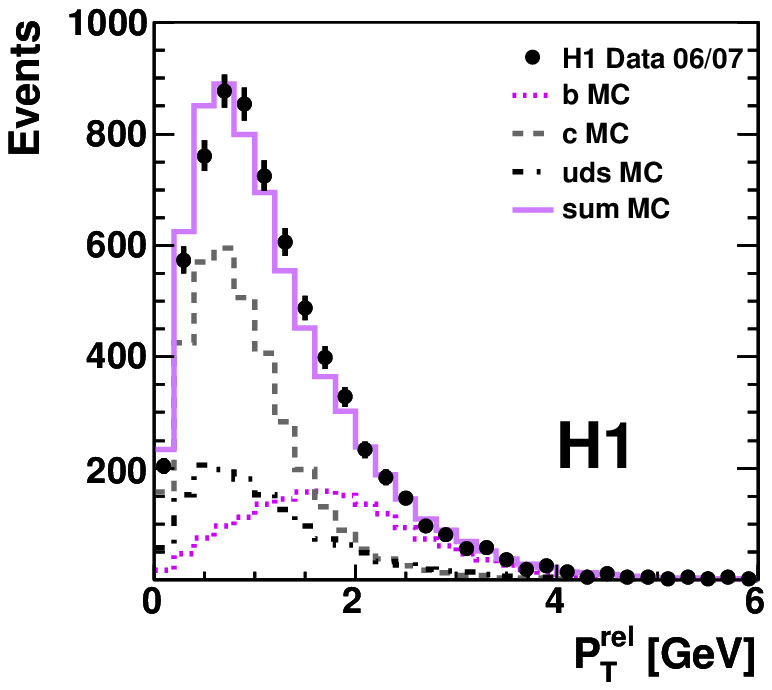,width = 9cm}}
\put(0.,7.2){(a)}
\put(9.2,7.2){(b)}
\end{picture}
\caption{The impact parameter \impactpar \ and \ptrel \ distributions for the total event sample. The data are compared to the \textsc{Pythia} predictions of the different quark contributions and their sum. The MC predictions are obtained from the fit to the total data sample.}
\label{fig:plotsflavourseparation}
\end{figure}

The two distributions that are used in the flavour separation are shown in figure \ref{fig:plotsflavourseparation}. The distribution of the impact parameter \impactpar \ shown in figure~\ref{fig:plotsflavourseparation} (a) is symmetric for $uds$ events while $b$ and $c$ events contribute more at large positive values of \impactpar. Therefore, the fit of this variable can distinguish the three different quark contributions. As shown in figure \ref{fig:plotsflavourseparation} (a), the sum of all three fitted contributions in the Monte Carlo simulation is able to describe the data quite well. This description is achieved by a better understanding of the detector and an improved detector simulation with regard to signal heights, noise levels and dead strips in the CST \cite{CSTTuning}, the inclusion of effects from alignment imperfections, and the description of the dead material in front of the CST and CJC. Therefore, a further smearing of measured track parameters in the simulation is not necessary, as it was done in the former H1 analysis using HERA I data \cite{H1Hera1desy05004}.
The \ptrel \ distribution is shown in figure \ref{fig:plotsflavourseparation} (b). The $uds$ and $c$ distributions are very similar and peak at low values of \ptrel, while the $b$ events contribute more at higher values of \ptrel. The sum of all quark contributions in the Monte Carlo simulation is able to describe the data reasonably well.
\par
The fitted parameters $f_{b}$, $f_{c}$ and $f_{l}$ for the whole kinematic range are:
\begin{align}\nonumber
f_b&=(26.0\pm1.2)\%,\\ \nonumber
f_c&=(48.6\pm2.5)\%,\\ \nonumber
f_l&=(25.3\pm2.6)\%. \nonumber
\end{align}
The $\chi^2/ndf$ is found to be $0.76$ for the total sample. 
\par
\begin{figure}
\setlength{\unitlength}{1cm}
\centering
\begin{picture}(15.,21.)(0.,0.)
\put(-1.5,15.){\epsfig{file=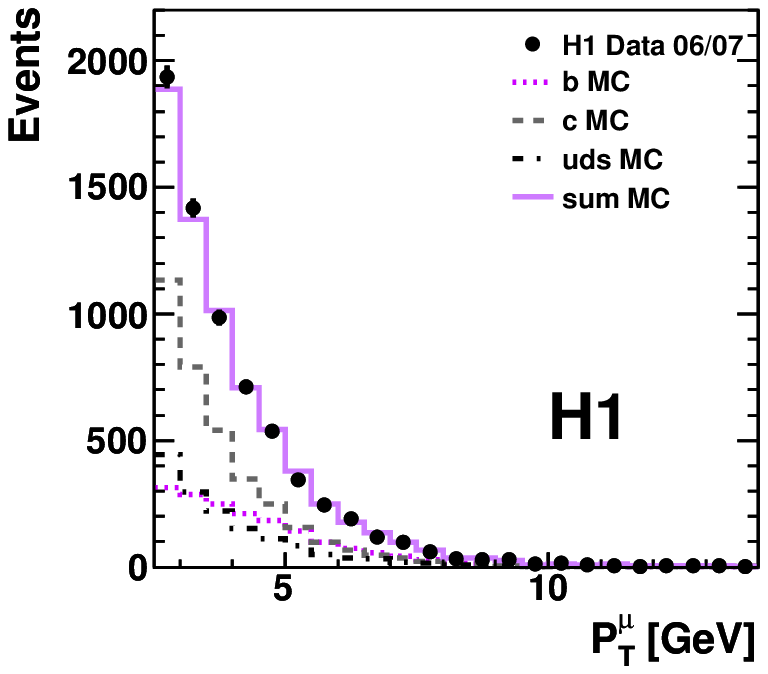,width = 9cm}}
\put(7.7,15.){\epsfig{file=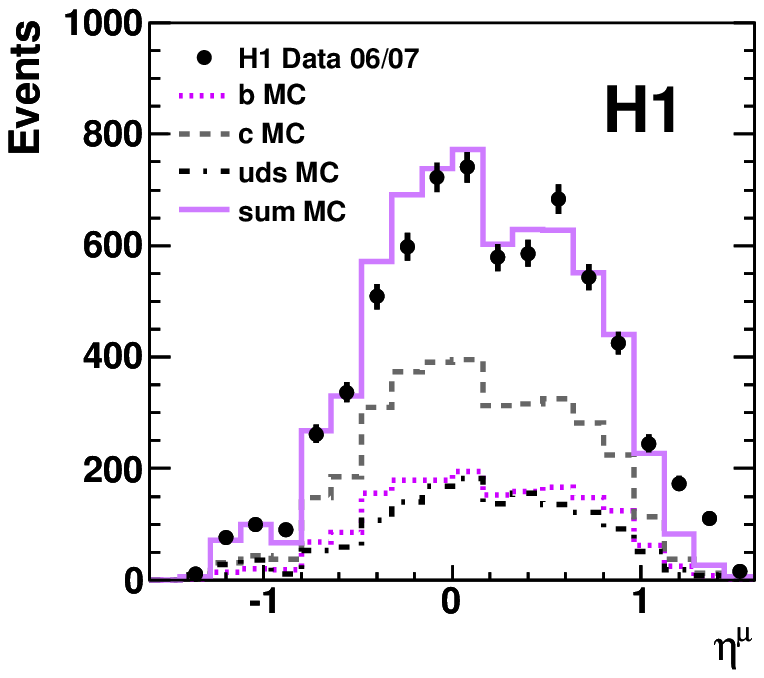,width=9cm}}
\put(-1.5,7.5){\epsfig{file=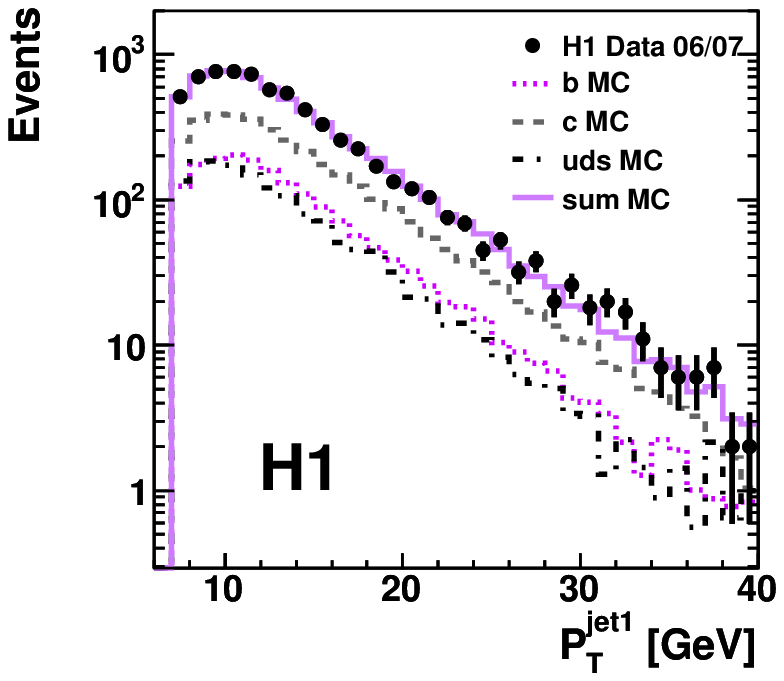,width = 9cm}}
\put(7.7,7.5){\epsfig{file=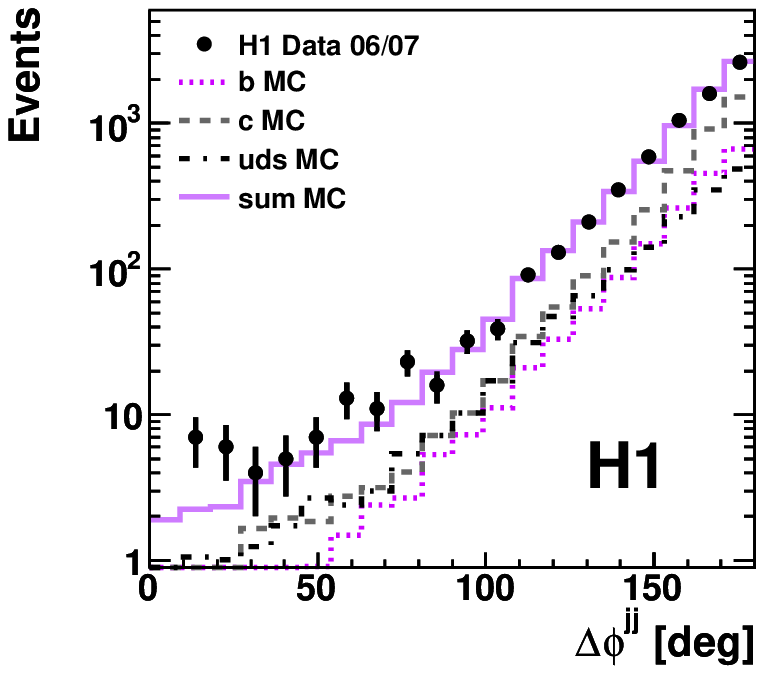,width = 9cm}}
\put(-1.5,0.){\epsfig{file=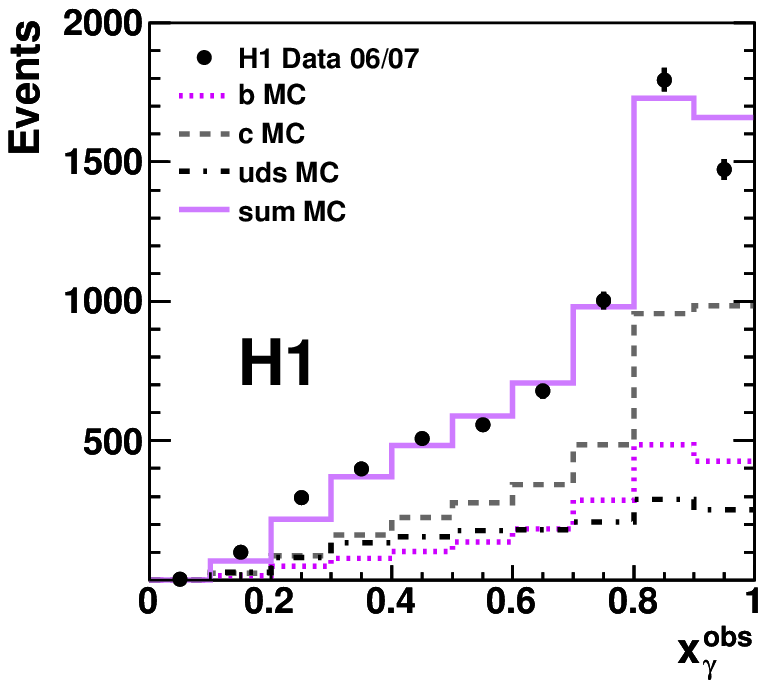,width = 9cm}}
\put(7.7,0.){\epsfig{file=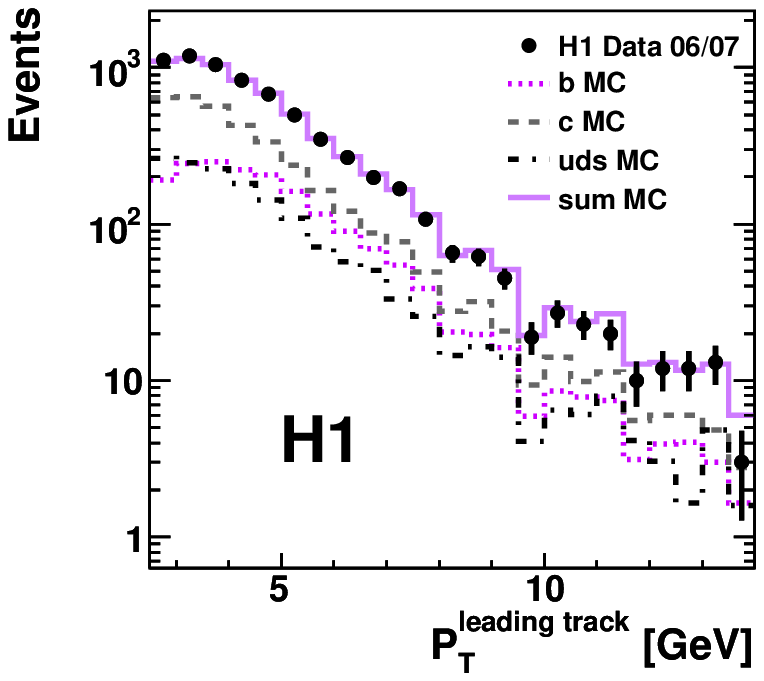,width = 9cm}}
\put(0.,22.1){(a)}
\put(9.2,22.1){(b)}
\put(0.,14.6){(c)}
\put(9.2,14.6){(d)}
\put(0.,7.2){(e)}
\put(9.2,7.2){(f)}
\end{picture}
\caption{Distributions of kinematic variables for the selected event sample. $P_T^{leading track}$ is the transverse momentum of the track with the highest transverse momentum in the event. The other notations are described in the text. The data are compared to the \textsc{Pythia} predictions of the different quark contributions and their sum. The MC predictions are obtained from the fit to the total data sample.}
\label{fig:controlplots}
\end{figure}

Control distributions for the data sample in comparison to the Monte Carlo simulations are shown in figure \ref{fig:controlplots}. All selection cuts are applied. The data are compared with the MC contributions from beauty, charm and light quark events and their sum with the relative fractions taken from the fit as discussed above. The number of events in the simulation is normalised to the one of the data. It is observed that the shapes of the MC contributions are rather similar for beauty, charm and light quark events for the distributions shown here. For the determination of the detector corrections a reweighting of \dphijets\  and the transverse momentum of the leading selected track in the event is performed on hadron level in the Monte Carlo simulation to provide a better description of the data. Only small deviations between data and MC are observed, such as in the forward \etamuon \ region in figure \ref{fig:controlplots} (b) and in the very small \dphijets \ region in \ref{fig:controlplots} (d). 

\subsection{Cross Section Determination}
Total and differential visible beauty and charm $ep$ cross sections are measured in the photoproduction regime. The fitted fractions $f_b$ and $f_c$ are converted to cross sections $\sigma_{b(c)}$ in each bin using

\begin{equation}
\sigma_{b(c)}=\frac{f_{b(c)}N_{Data}N_{b(c)}^{MCgen}}{\mathcal{L}N_{b(c)}^{MCrec}}.
\end{equation}
Here, $N_{Data}$ and $N_{b(c)}^{MCrec}$ represent the number of data or Monte Carlo simulation events passing all selection cuts on reconstruction level. The variable $N_{b(c)}^{MCgen}$ stands for the events selected in the Monte Carlo simulation on hadron level and $\mathcal{L}$ denotes the luminosity of the data. The differential cross sections are obtained by dividing by the bin width.

%%%%%%%%%%%%%%%%%%%%%%%%%%%%%%%%%%%%%%%%%%%%%%%%%%%%%%%%%%
\section{Fake Muon Rate}
%%%%%%%%%%%%%%%%%%%%%%%%%%%%%%%%%%%%%%%%%%%%%%%%%%%%%%%%%%
All backgrounds to semi-muonic $b$ and $c$ decays are called {\it fake muons} here. 
These contributions are modelled using MC simulation and originate mainly
from $uds$ events, with a small fraction from $b$ and $c$ events.  
Three sources of fake muons are considered:
\begin{itemize}
\item{Hadrons which reach the muon detector and are misidentified as muons. According to the fully inclusive \textsc{Pythia} MC, $1.6\%$ of the selected $b$ events and $2.3\%$ of the selected $c$ events originate from hadron misidentification. } 
\item{Muons which do not originate from $b$ or $c$ hadron decays, but from other hadrons such as kaons and pions. This background source is denoted as {\it inflight decay} in the following. 
According to the full inclusive \textsc{Pythia} MC, $0.9\%$ of the $b$ and $0.7\%$ of the selected $c$ events originate from inflight decays.}
\item{Cosmic ray muons which coincide in time with real $ep$ events. About $1\%$ of the selected muon candidates are 
rejected as cosmic ray muons based on timing information from the CTD \cite{MiraKraemerThesis}. The remaining background from cosmic ray muons is negligible.}
\end{itemize}
The probability to fake a muon depends on the particle species.  The kinematic distributions of different particle species differ, but the $uds$ MC is used as one single template. Therefore the fake probabilities for the most important particle species are studied in data and MC.
The fake muon probabilities from misidentification and inflight decays $P^{\mu}_{h}$ for hadron $h$ are defined as: 
\begin{equation}\nonumber
P^{\mu}_h=\frac{\# \ \text{fake muons}}{\# \ \text{all hadrons}}.
\end{equation}
The fake muon probability is investigated for pions originating from $K^0_S\rightarrow \pi^+\pi^-$ decays and amounts to $P^{\mu}_{\pi}\approx0.05$ in the data. For protons the decay channels $\Lambda\rightarrow p\pi^-$ and $\Lambda\rightarrow \bar{p}\pi^+$ are used and the fake muon probability is found to be $P^{\mu}_{p}\approx0.04$ in the data. For $K^{\pm}$ mesons from the decay $D^{*\pm}\rightarrow D^0\pi_{slow}^{\pm}\rightarrow (K^{\mp}\pi^{\pm})\pi^{\pm}_{slow}$ a fake muon probability of $P^{\mu}_{K{\pm}}\approx0.01$ is measured in the data. It is observed that the pion and proton fake muon probabilities in the data are not described by the Monte Carlo simulation. They are reweighted in the Monte Carlo simulation by factors of $2.0$ and $1.9$, respectively, to match the data. The $K^{\pm}$ fake muon probability in the data and the simulation are in agreement. The misidentified muon events remain in the event sample.

%%%%%%%%%%%%%%%%%%%%%%%%%%%%%%%%%%%%%%%%%%%%%%%%%%%%%%%%%%%
\section{Systematic Uncertainties}
%%%%%%%%%%%%%%%%%%%%%%%%%%%%%%%%%%%%%%%%%%%%%%%%%%%%%%%%%%
The following uncertainties are taken into account in order to evaluate the systematic errors.  
\begin{itemize}
\item{The trigger efficiencies are determined using independent trigger channels in the DIS regime since no independent triggers exist in photoproduction. The uncertainty is estimated by the difference between the efficiency found in the data and the simulation. The Monte Carlo simulation is reweighted to match the data trigger efficiency.}
\item{The efficiency for the identification of the muons is determined using a high statistics sample of events of elastically produced $J/\psi$ mesons \cite{MartinGoettlichThesis}. The efficiency is known to a precision of $4\%$.}
\item{The track efficiency of the CTD is known to $\pm1\%$ and that of the CST to $\pm2\%$. The uncertainty due to the track efficiencies is estimated by varying the efficiencies of the CTD and CST correspondingly.}
\item{The integrated luminosity is known to a precision of $4\%$.}
\item{The uncertainty due to the resolution of the impact parameter $\delta$ of the muon tracks is estimated by varying the resolution by an amount that encompasses any difference between the data and the simulation. This is achieved by applying an additional Gaussian smearing in the Monte Carlo simulation of $200 \ \mu\text{m}$ to $5\%$ of randomly selected tracks and $12\ \mu\text{m}$ to the rest.}
\item{The uncertainty on the cross section arising from the uncertainty on the reconstruction of $\phi_{jet}$ is estimated by shifting its value by $\pm2^{\circ}$. }
\item{The uncertainty arising from the hadronic energy scale is estimated by changing it by $\pm1\%$ for the complete hadronic final state.}
\item{The dependence of the measurement on the physics model used for the templates representing different QCD evolution schemes is estimated by replacing the \textsc{Pythia} $b$ and $c$ Monte Carlo templates with \textsc{Cascade}.}
\item{The uncertainty on the cross section arising from the uncertainty of the parton fragmentation model is estimated by replacing the \textsc{Pythia} Monte Carlo samples using the Lund fragmentation function with samples based on the Peterson fragmentation function. }
\item{The uncertainty arising from fake muon background is estimated by not applying the weights that have been found in the fake rate probabilities for $K$ and $\Lambda$ decays.}
\item{The impact of the reweighting on the cross sections are investigated and found to be negligible.}

\end{itemize}
\par
The effect of the listed experimental uncertainties are estimated by varying the relevant variables in the Monte Carlo simulation or by modifying the corresponding efficiencies in the cross section calculation. The difference between the obtained cross sections with and without the change result in the measurement systematic uncertainties, which are summarised in table \ref{table_syserr}.
The individual effects of the above experimental uncertainties are combined in quadrature, yielding a total uncertainty of $10.5\%$ and $10.4\%$ on the measured $b$ and $c$ cross sections, respectively. The systematic uncertainties as derived from the integrated sample are applied to each analysis bin in order to avoid statistical fluctuations. The largest contribution to this uncertainty for the $c$ measurement arises from systematics attributed to the hadronic energy scale ($5\%$). The systematic errors of the $b$ analysis are not dominated by a single source. 
\begin{table}
\centering
\begin{tabular}{|l|c|c|}
 \hline
 Systematic error source & Beauty $\Delta\sigma/\sigma[\%]$& Charm $\Delta\sigma/\sigma[\%]$\\ \hline 
 Trigger efficiency &  $4$ & $4$ \\ 
 Muon identification & $4$ & $4$ \\
 Track finding efficiency & $3$ & $3$\\
 Luminosity & $4$ & $4$ \\ 
 $\delta$ Resolution & $3$ & $2$  \\ 
Jet axis & $4$  & $2$ \\
 Hadronic energy scale & $3$ & $5$ \\  
 Physics model &  $3$ &  $1$ \\
 Fragmentation &  $3$ & $4$ \\
 Fake muon background & $1$ & $1$ \\ \hline \hline
Total & $10.5$  & $10.4$ \\  \hline
\end{tabular}
\caption{Summary of the systematic uncertainties of the beauty and charm cross sections.}\label{table_syserr}
\end{table}

%%%%%%%%%%%%%%%%%%%%%%%%%%%%%%%%%%%%%%%%%%%%%%%%%%%%%%%%%%
\section{Results}
%%%%%%%%%%%%%%%%%%%%%%%%%%%%%%%%%%%%%%%%%%%%%%%%%%%%%%%%%%
The cross sections for $b$ and $c$ in photoproduction using semi-muonic decays in dijet events are measured. The cross sections are determined for the phase space defined by the kinematic range and the event selection cuts presented in table \ref{CutSummaryTable}. The measured cross sections are compared to the expectations of the MC programs {\sc Pythia}, {\sc Cascade}, {\sc Herwig}, and MC@NLO. The total measured and predicted cross sections are listed in table \ref{table_totalresults}. 
{\sc Pythia} shows the highest normalisation of the three LO MCs, while the normalisation of {\sc Cascade} is below the one of {\sc Pythia} and {\sc Herwig} has the lowest normalisation. For the beauty measurement, the {\sc Pythia} prediction is closest to the data and gives the best description of the three LO MC predictions.
The beauty and charm data cross sections tend to be underestimated by the MC@NLO predictions but are in agreement within the errors. The precision of the measured cross sections are much higher than the ones of the theory predictions shown here.
\par
\begin{figure}
\setlength{\unitlength}{1cm}
\centering
\begin{picture}(15.,21.)(0.,0.)
\put(-1.5,15.){\epsfig{file=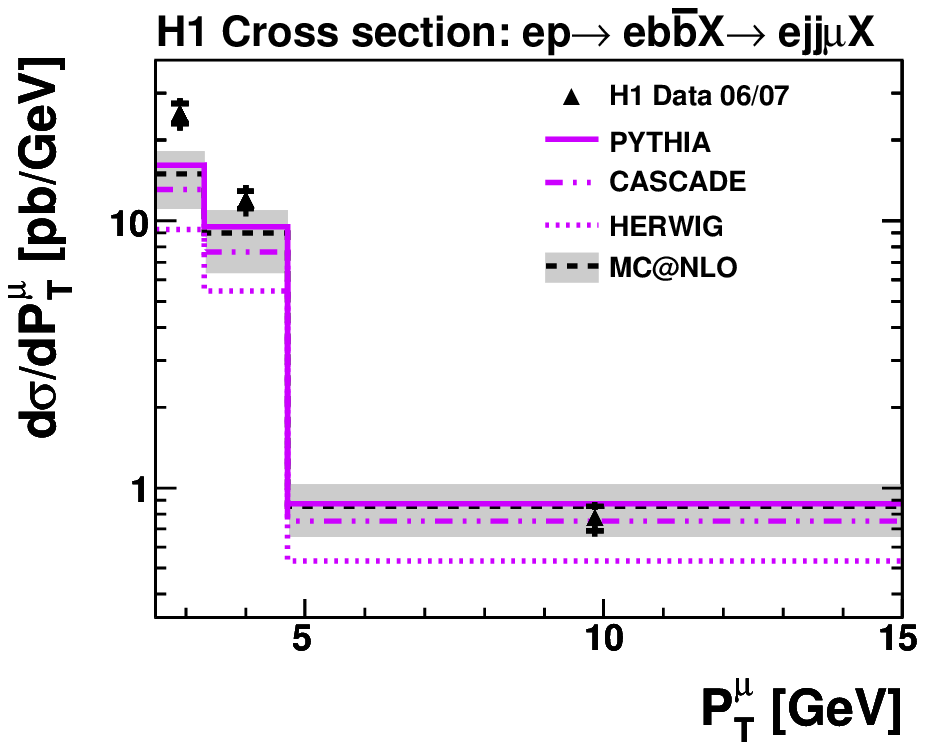,width = 8.5cm}}
\put(7.7,15.){\epsfig{file=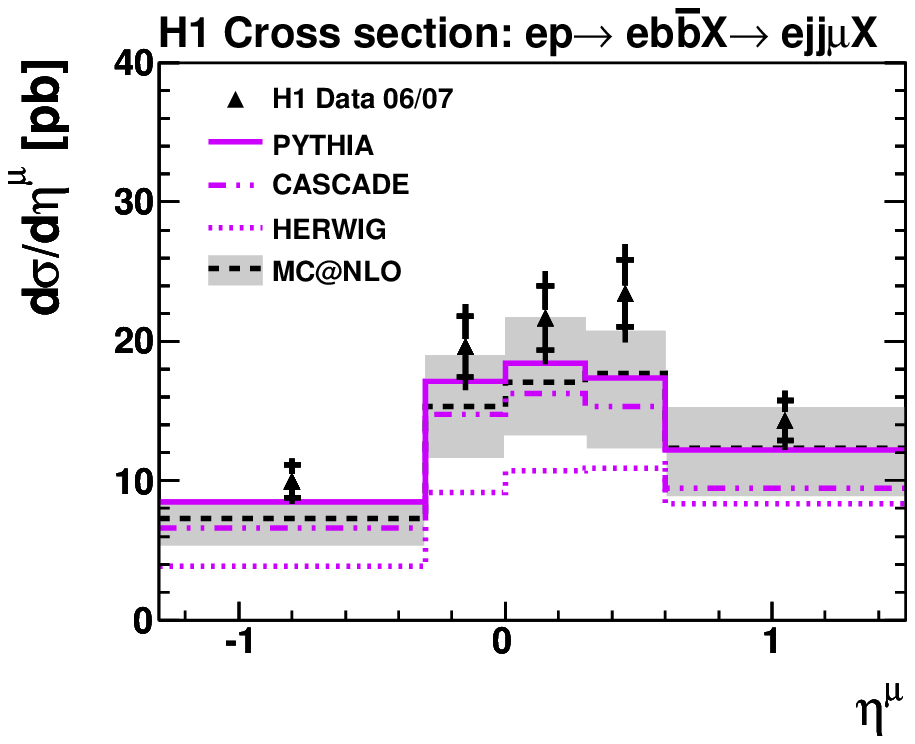,width=8.5cm}}
\put(-1.5,7.5){\epsfig{file=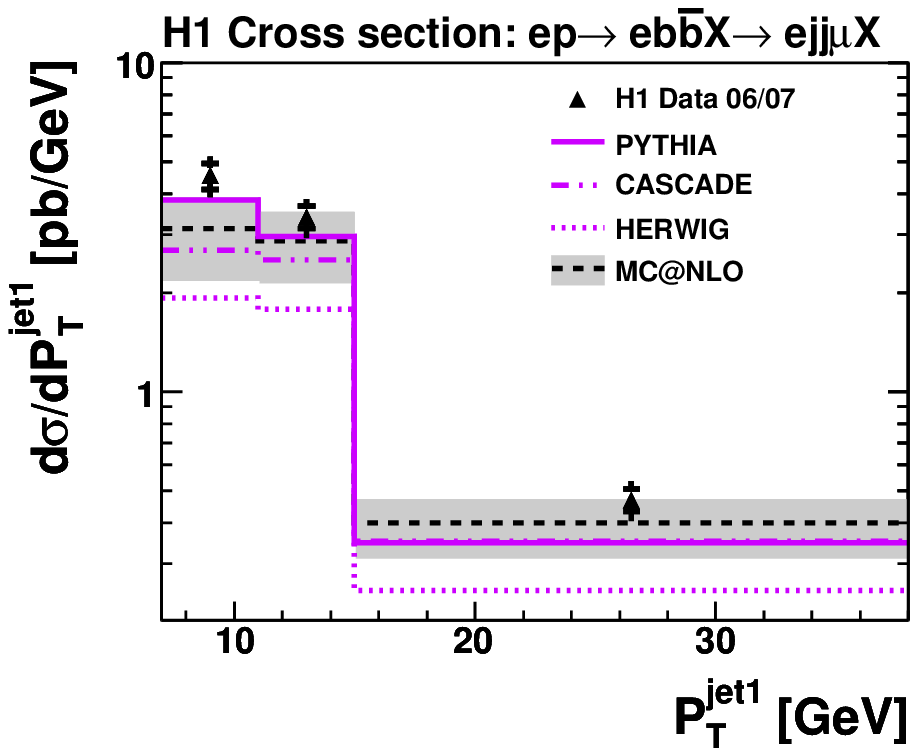,width = 8.5cm}}
\put(7.7,7.5){\epsfig{file=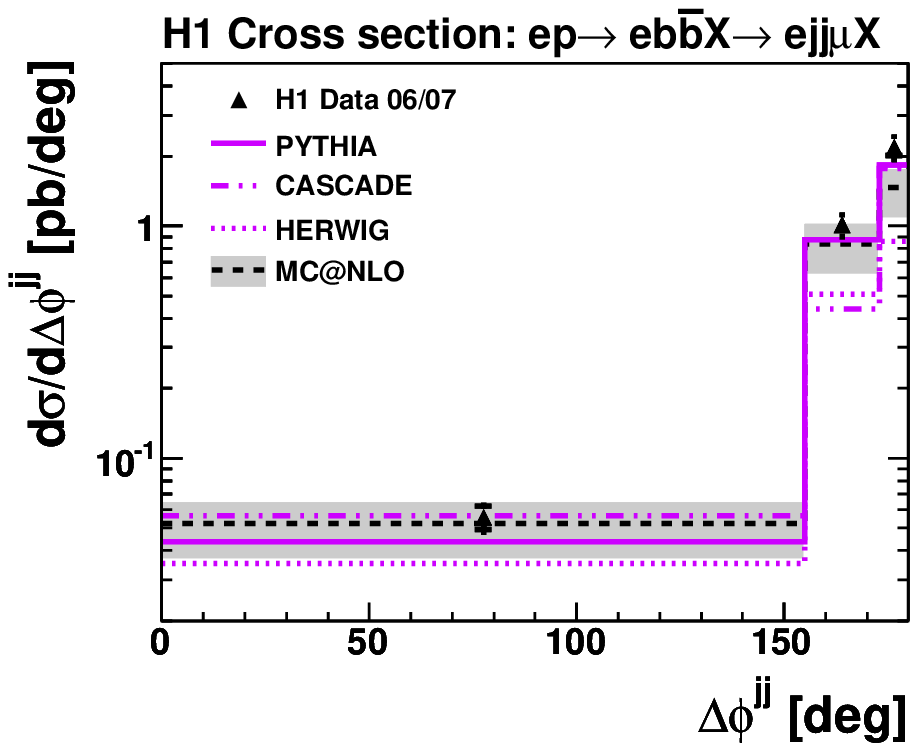,width = 8.5cm}}
\put(-1.5,0.){\epsfig{file=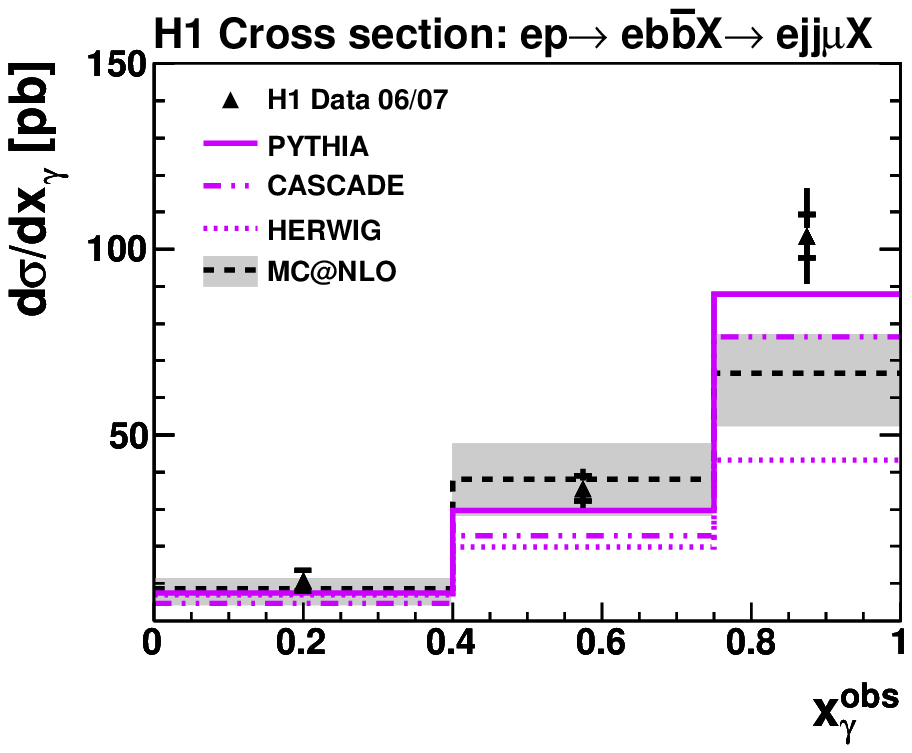,width = 8.5cm}}
\put(0.,22.1){(a)}
\put(9.2,22.1){(b)}
\put(0.,14.6){(c)}
\put(9.2,14.6){(d)}
\put(0.,7.2){(e)}
\end{picture}
\caption{The differential cross sections for beauty photoproduction of dijet events with a muon as a function of \ptmuon, \etamuon, \ptjetone, \dphijets, and \xgamma. The inner error bars show the statistical errors, the outer error bars represent the statistical and systematic errors added in quadrature. The data are compared with the predictions from the LO models \textsc{Pythia},  \textsc{Cascade} and \textsc{Herwig} as well as to the NLO predictions of MC@NLO. The theoretical uncertainties of MC@NLO are given as shaded band.}
\label{fig:beautyxsecfullsample}

\end{figure}
\begin{figure}
\setlength{\unitlength}{1cm}
\centering
\begin{picture}(15.,15.)(0.,0.)
\put(-1.5,7.5){\epsfig{file=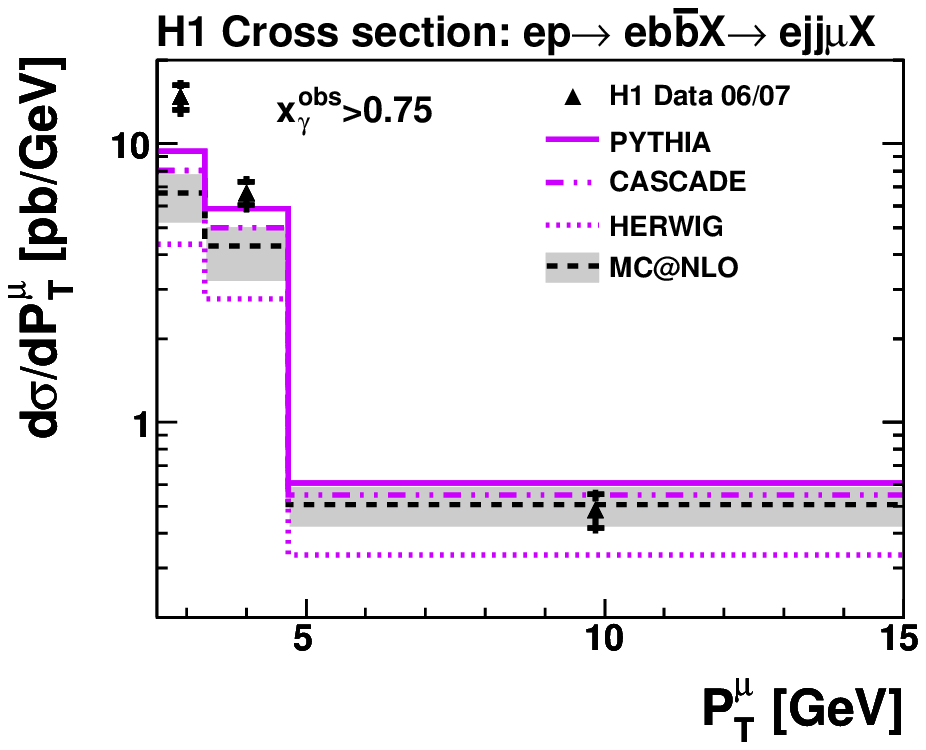,width = 8.5cm}}
\put(7.7,7.5){\epsfig{file=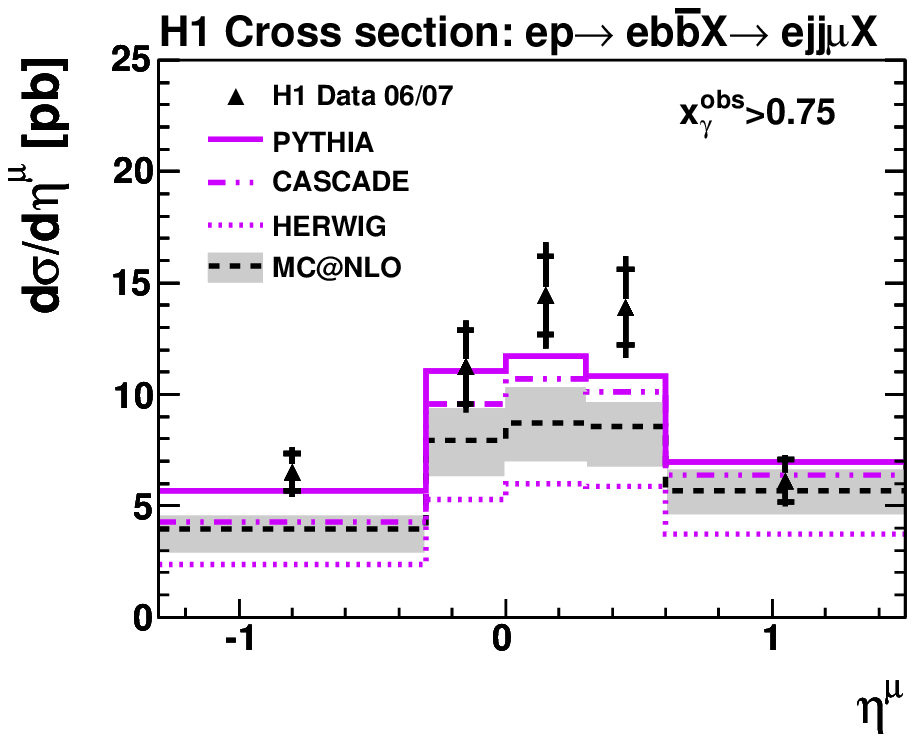,width=8.5cm}}
\put(-1.5,0.){\epsfig{file=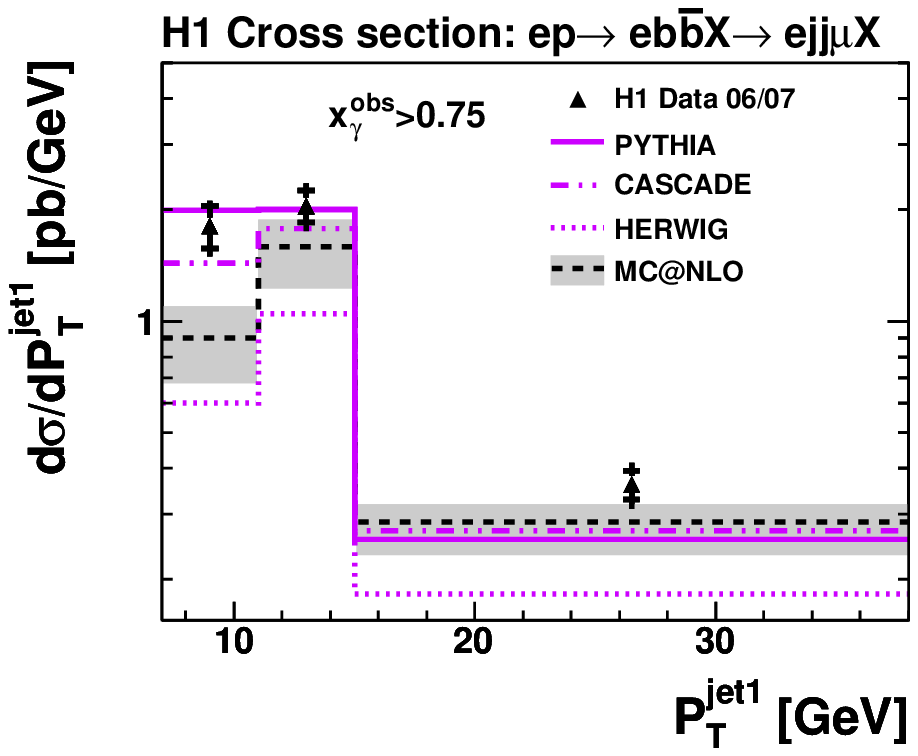,width = 8.5cm}}
\put(7.7,0.){\epsfig{file=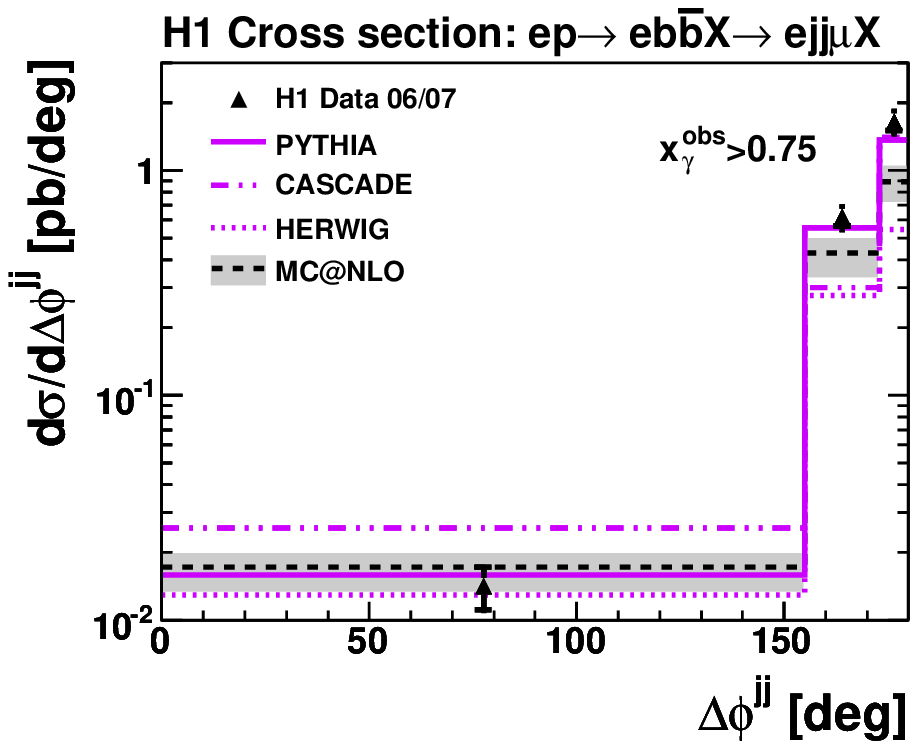,width = 8.5cm}}
\put(0.,14.6){(a)}
\put(9.2,14.6){(b)}
\put(0.,7.2){(c)}
\put(9.2,7.2){(d)}
\end{picture}
\caption{The differential cross sections for beauty photoproduction of dijet events using semi-muonic decays for $\xgamma>0.75$ as a function of \ptmuon, \etamuon, \ptjetone, and \dphijets. For details see caption of figure~\ref{fig:beautyxsecfullsample}.}
\label{fig:beautyxsecdirectsample}
\end{figure}
\begin{figure}
\setlength{\unitlength}{1cm}
\centering
\begin{picture}(15.,15.)(0.,0.)
\put(-1.5,7.5){\epsfig{file=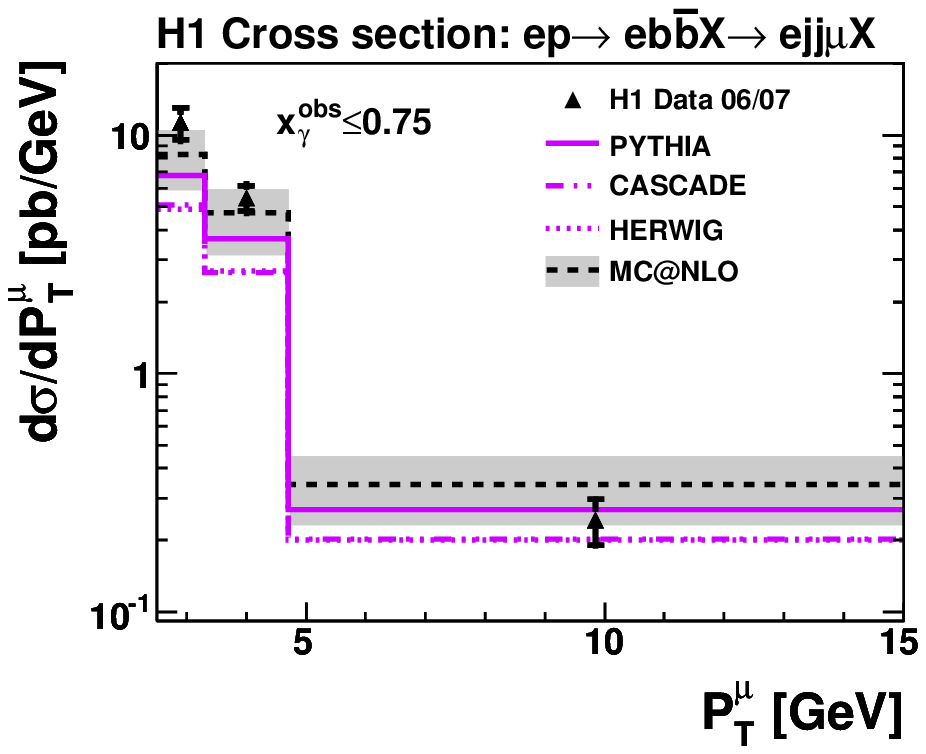,width = 8.5cm}}
\put(7.7,7.5){\epsfig{file=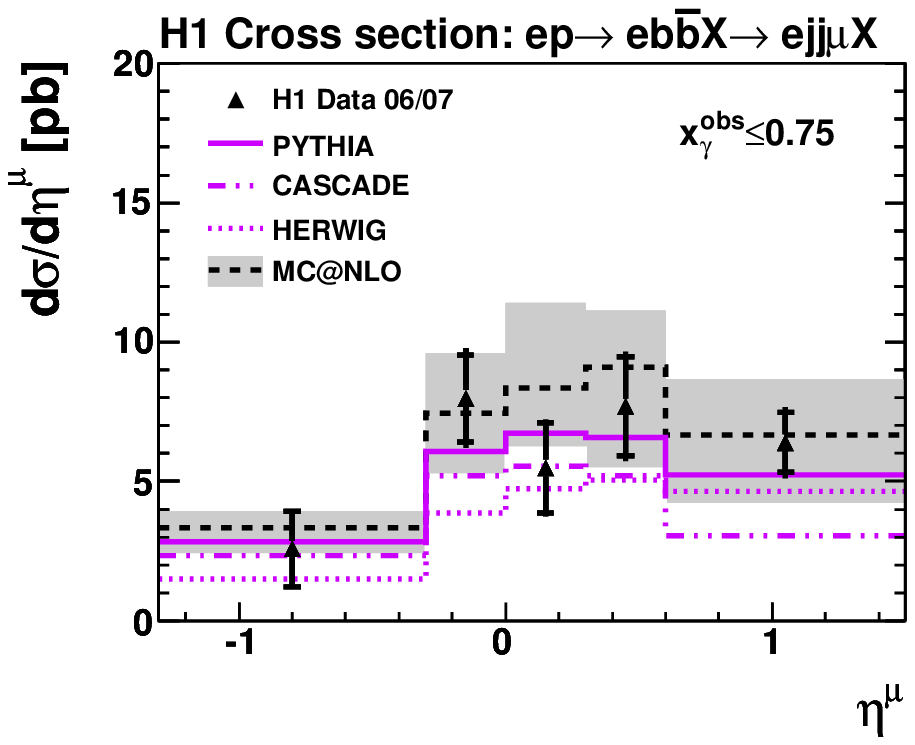,width=8.5cm}}
\put(-1.5,0.){\epsfig{file=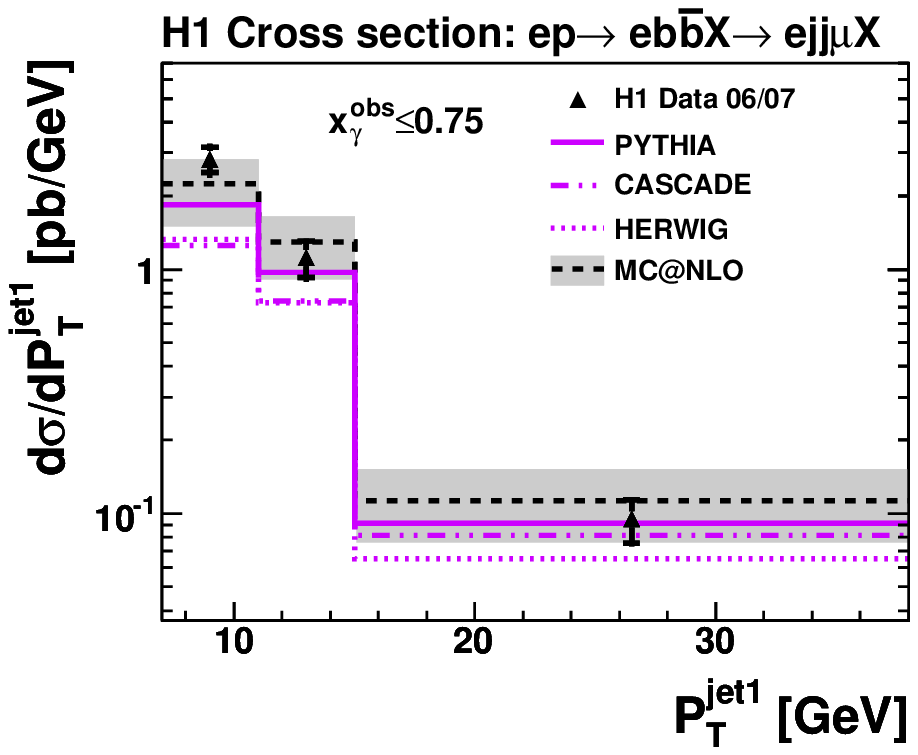,width = 8.5cm}}
\put(7.7,0.){\epsfig{file=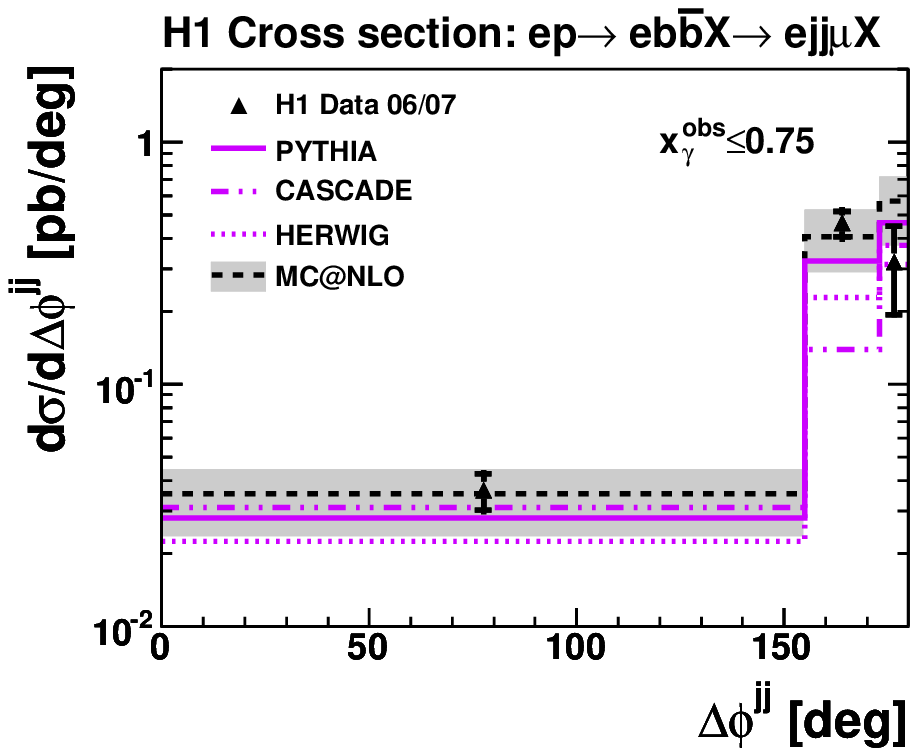,width = 8.5cm}}
\put(0.,14.6){(a)}
\put(9.2,14.6){(b)}
\put(0.,7.2){(c)}
\put(9.2,7.2){(d)}
\end{picture}
\caption{The differential cross sections for beauty photoproduction of dijet events using semi-muonic decays for $\xgamma\leq0.75$ as a function of \ptmuon, \etamuon, \ptjetone, and \dphijets. For details see caption of figure~\ref{fig:beautyxsecfullsample}.}
\label{fig:beautyxsecresolvedsample}
\end{figure}

The beauty and charm cross sections are measured differentially as a function of the transverse momentum of the leading jet \ptjetone\ and of the muon \ptmuon, the pseudorapidity of the muon \etamuon, the momentum fraction \xgamma\  carried by the photon entering the hard interaction and the azimuthal angular difference \dphijets\  between the two leading jets. The measurements are performed for the full sample, as well as for direct and resolved enriched processes separately. The distinction is performed by the variable $\xgamma$, which leads to enriched resolved processes in the region $\xgamma\leq0.75$ and direct photon enriched processes for $\xgamma>0.75$. The cross sections for the beauty measurements are given in tables \ref{tab:beautyfull}-\ref{tab:beautyresolved} and shown in figures \ref{fig:beautyxsecfullsample}-\ref{fig:beautyxsecresolvedsample}. 
\begin{table}
\centering
\begin{tabular}{|c|c|c|}
\hline
                   & $\sigma_{\text{vis}}(ep\rightarrow eb\bar{b}X\rightarrow ejj\mu X')$ [pb]  & $\sigma_{\text{vis}}(ep\rightarrow ec\bar{c}X\rightarrow ejj\mu X')$ [pb]  \\ \hline  \hline
H1 Data  & $43.3\pm 2.1\ (\text{stat.})\pm 4.5\ (\text{sys.})$ & $81.3\pm 4.3\ (\text{stat.})\pm 8.5\ (\text{sys.})$ \\ \hline
\textsc{Pythia} & $35.3$  & $94.3$ \\
\textsc{Cascade} &  $29.0$ & $76.8$ \\
\textsc{Herwig}  & $20.6$ & $58.5$ \\
MC@NLO &  $33.4^{+7.1}_{-9.2}$   &  $58.6^{+29.5}_{-11.2}$   \\ \hline
\end{tabular}
\caption{Total visible measured beauty and charm cross sections along with their statistical and systematic errors. The total predictions from \textsc{Pythia}, \textsc{Cascade}, \textsc{Herwig}, and MC@NLO are also shown. The MC@NLO predictions are given with their theoretical uncertainties. }\label{table_totalresults}
\end{table}
\par
In the case of beauty production the models provide a good description of the measured cross sections in terms of shape in all distributions.  For the LO MCs this is true for the full sample, as well as for the direct and resolved enriched regions. 
The cross sections for direct enriched processes are well described in shape, but tend to be underestimated by MC@NLO, while for resolved enriched processes a reasonable agreement is observed both in shape and normalisation.
\par
In the analysis of semi-muonic $b$ decays in dijet events with HERA I data \cite{H1Hera1desy05004} an excess of data compared to the NLO predictions of the FMNR program was observed in the first bin of \ptmuon\ and \ptjetone. In this analysis the NLO predictions are provided by MC@NLO which is based on the FMNR parton level calculations. 
%The predictions from FMNR are found to agree with those of MC@NLO within the theoretical uncertainties. The excess in the first bin of  \ptmuon\ and \ptjetone, which was seen in the HERA I analysis, is not confirmed. Especially the first bin of \ptjetone\ does not show such an excess in the full sample. 
Also in this analysis, the NLO predictions lie below the data in the
first bin of \ptmuon\ and \ptjetone, but they are consistent with the data
within $2 \sigma$ of the experimental and theoretical uncertainty.
Whereas in enriched direct processes, the data tend to be underestimated by the MC@NLO prediction in the first bin of \ptmuon \ and the first bin of \ptjetone, in enriched resolved processes, no such effect is visible. 
\par
The cross sections as a function \dphijets \ show a significant contribution away from the back-to-back configuration at \dphijets$\simeq 180^{\circ}$. Such a configuration can be described by models which include significant contributions from higher order QCD radiation or a transverse momentum of the gluon in the initial state. This distribution is reasonably well described by all models. In direct and resolved enriched processes, this observation also holds.  
\par
\begin{figure}
\setlength{\unitlength}{1cm}
\centering
\begin{picture}(15.,21.)(0.,0.)
\put(-1.5,15.){\epsfig{file=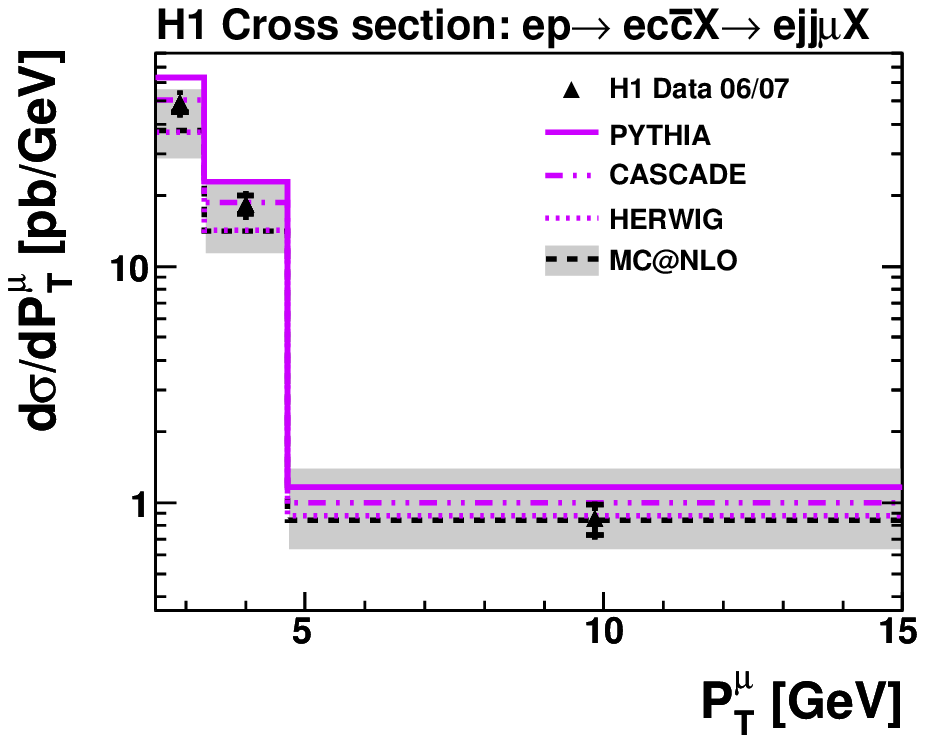,width = 8.5cm}}
\put(7.7,15.){\epsfig{file=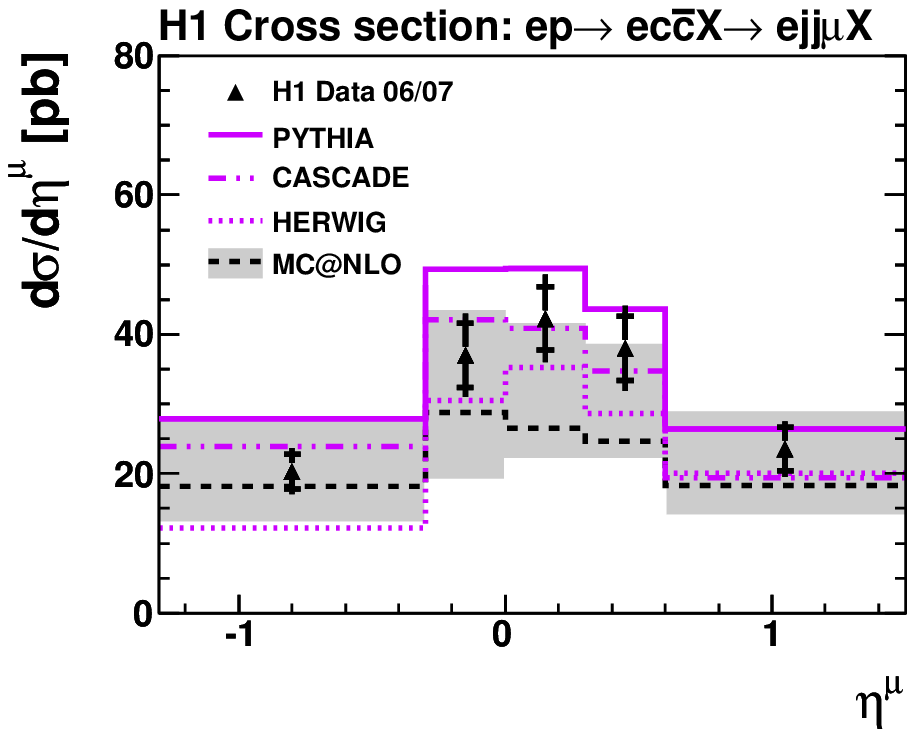,width=8.5cm}}
\put(-1.5,7.5){\epsfig{file=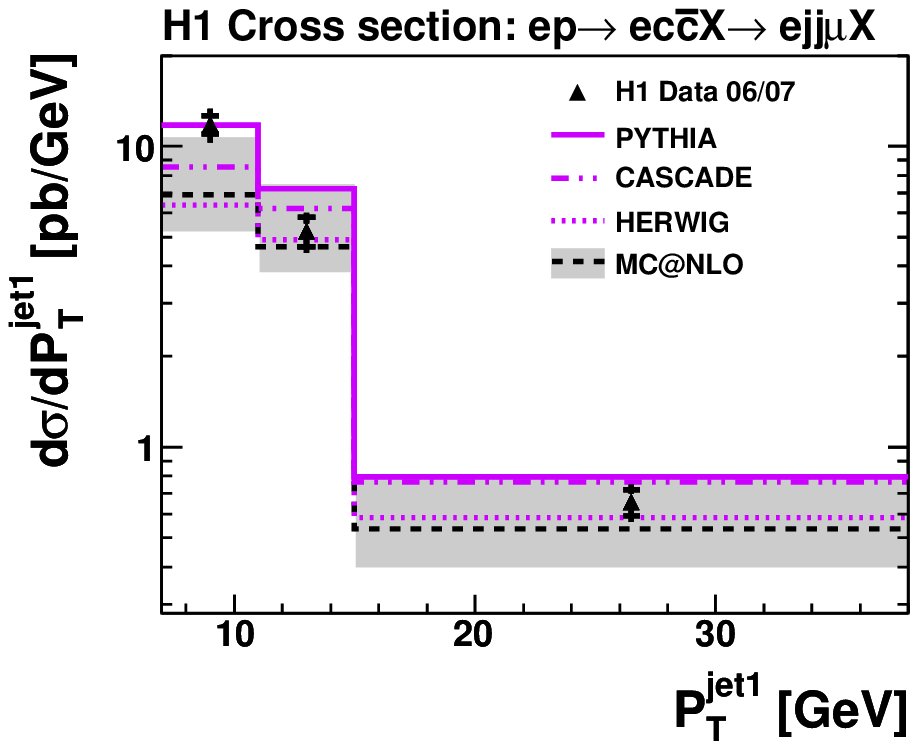,width = 8.5cm}}
\put(7.7,7.5){\epsfig{file=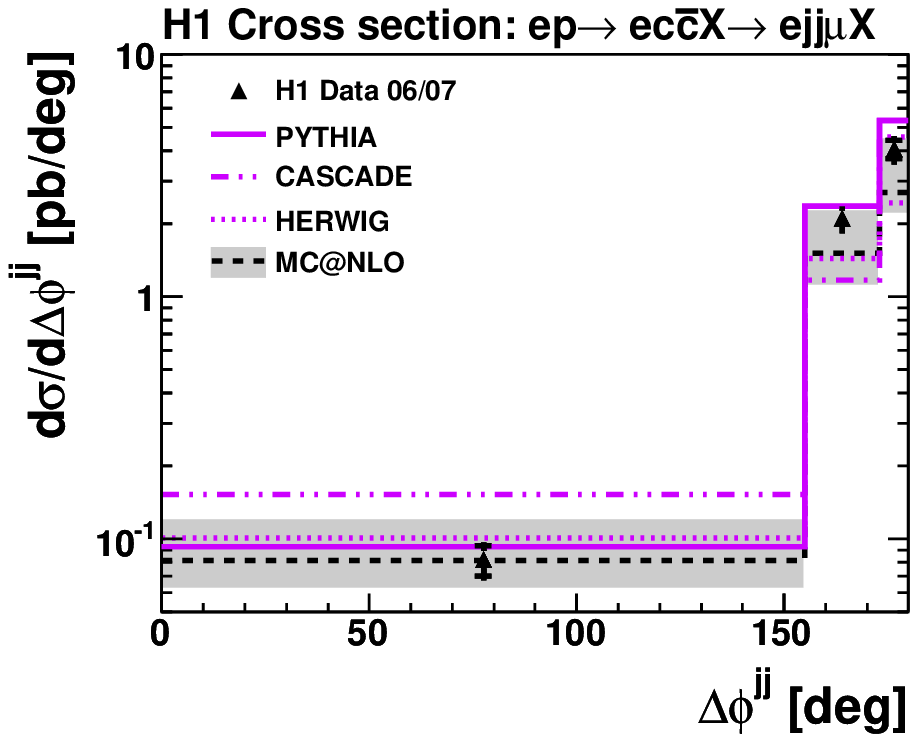,width = 8.5cm}}
\put(-1.5,0.){\epsfig{file=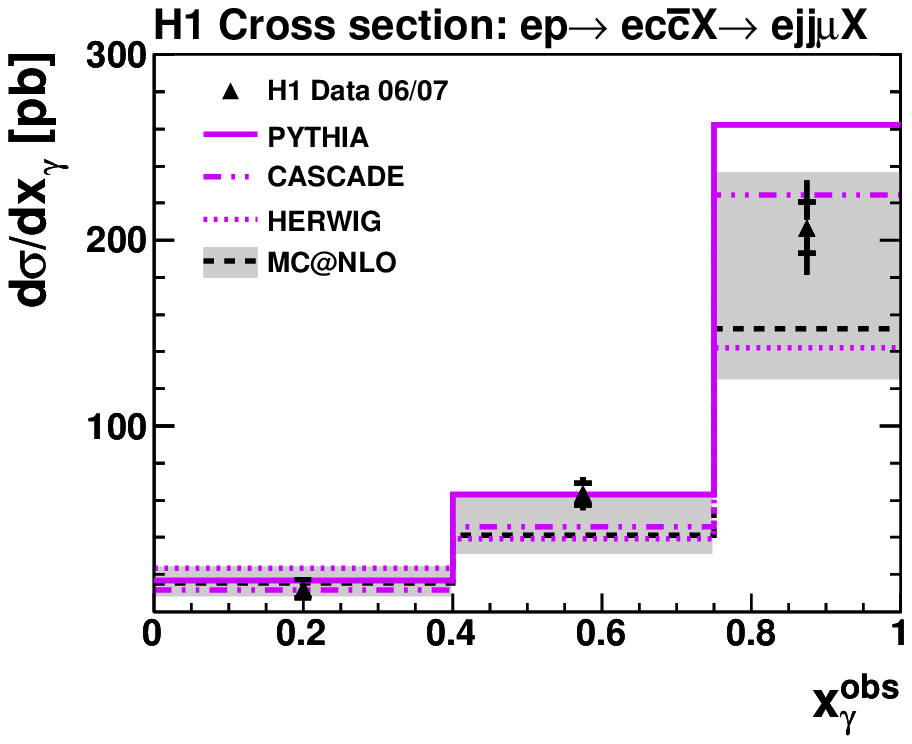,width = 8.5cm}}
\put(0.,22.1){(a)}
\put(9.2,22.1){(b)}
\put(0.,14.6){(c)}
\put(9.2,14.6){(d)}
\put(0.,7.2){(e)}
\end{picture}
\caption{The differential cross sections for charm photoproduction of dijet events using semi-muonic decays as a function of \ptmuon, \etamuon, \ptjetone, \dphijets, and \xgamma. For details see caption of figure~\ref{fig:beautyxsecfullsample}.}\label{fig:charmxsecfullsample}
\end{figure}
\begin{figure}
\setlength{\unitlength}{1cm}
\centering
\begin{picture}(15.,15.)(0.,0.)
\put(-1.5,7.5){\epsfig{file=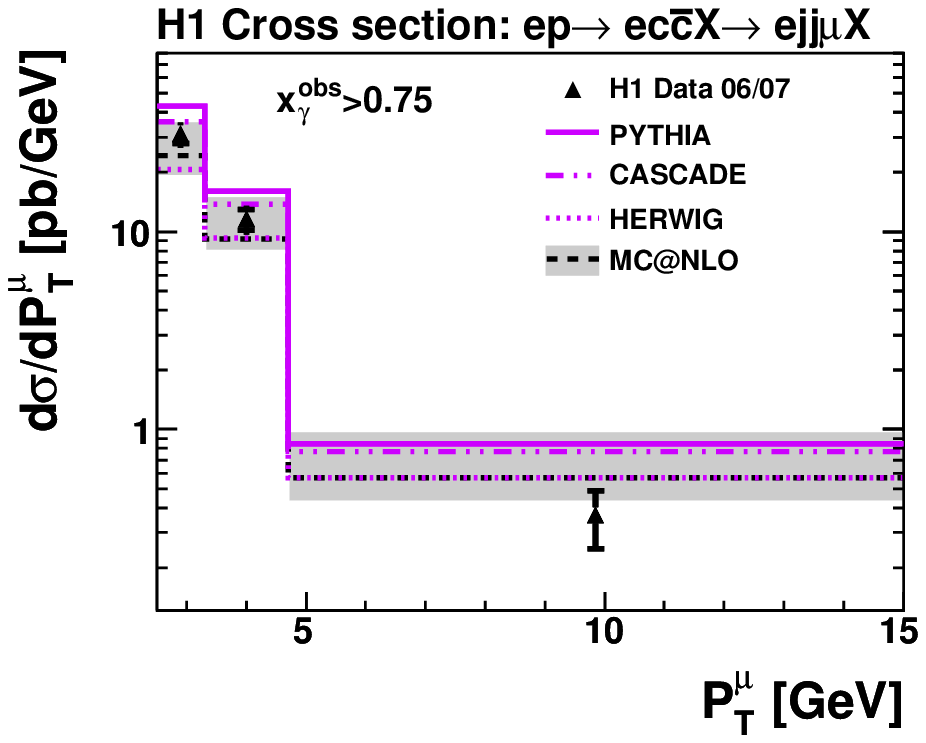,width = 8.5cm}}
\put(7.7,7.5){\epsfig{file=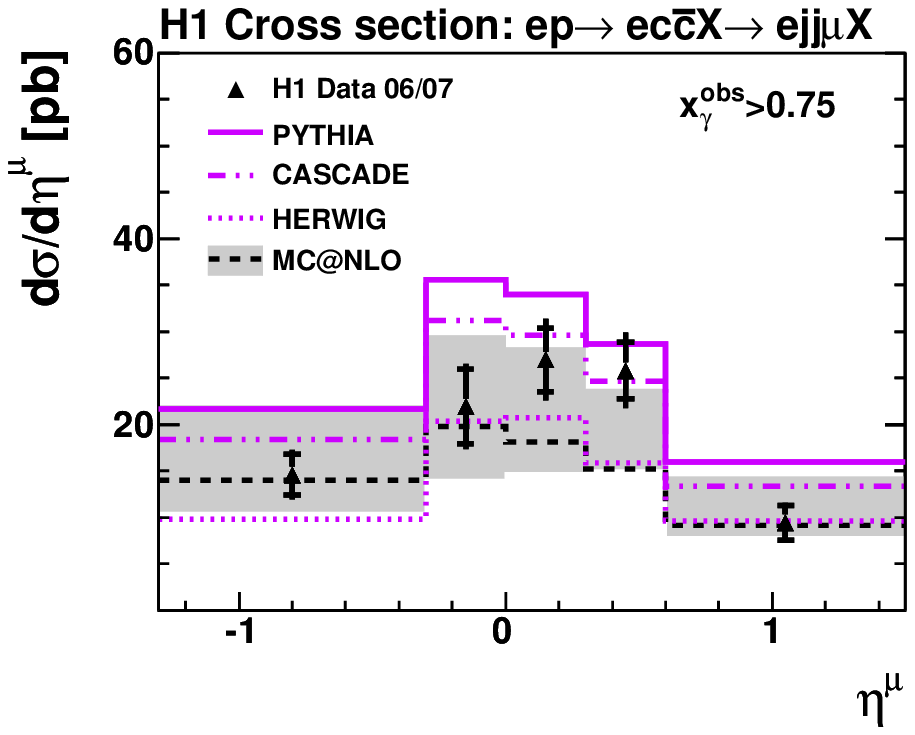,width=8.5cm}}
\put(-1.5,0.){\epsfig{file=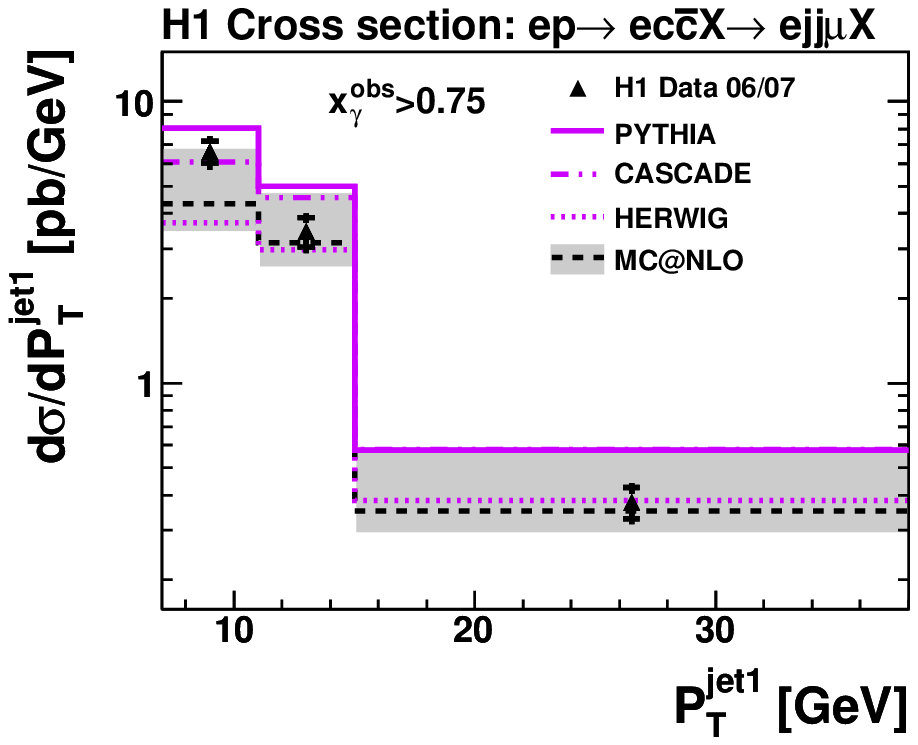,width = 8.5cm}}
\put(7.7,0.){\epsfig{file=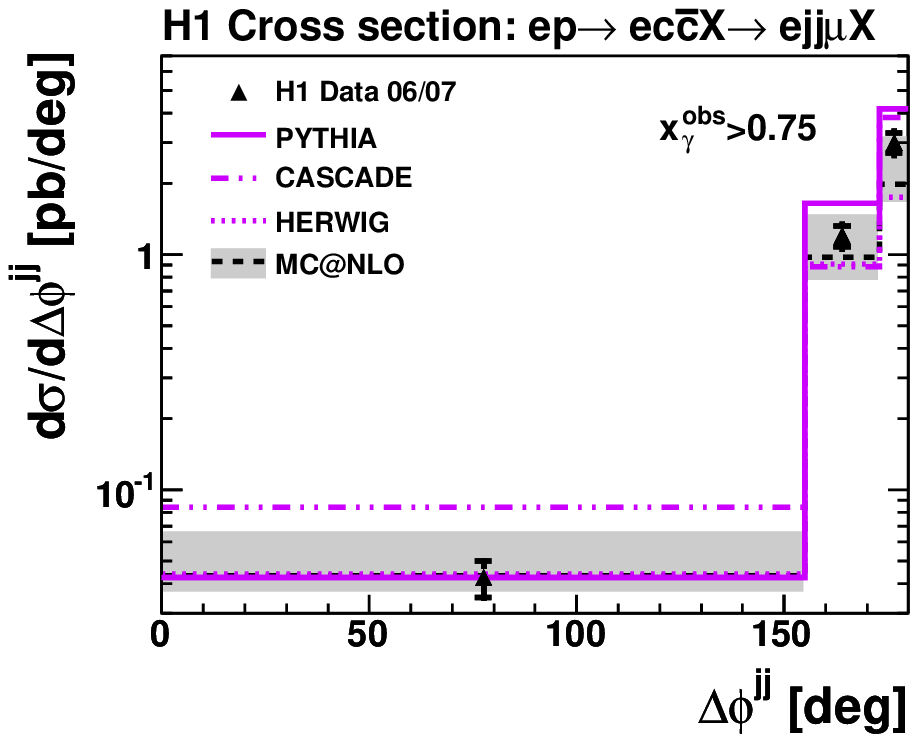,width = 8.5cm}}
\put(0.,14.6){(a)}
\put(9.2,14.6){(b)}
\put(0.,7.2){(c)}
\put(9.2,7.2){(d)}
\end{picture}
\caption{The differential cross sections for charm photoproduction of dijet events using semi-muonic decays for $\xgamma>0.75$ as a function of \ptmuon, \etamuon, \ptjetone, and \dphijets. For details see caption of figure~\ref{fig:beautyxsecfullsample}.}
\label{fig:charmxsecdirectsample}
\end{figure}
\begin{figure}
\setlength{\unitlength}{1cm}
\centering
\begin{picture}(15.,15.)(0.,0.)
\put(-1.5,7.5){\epsfig{file=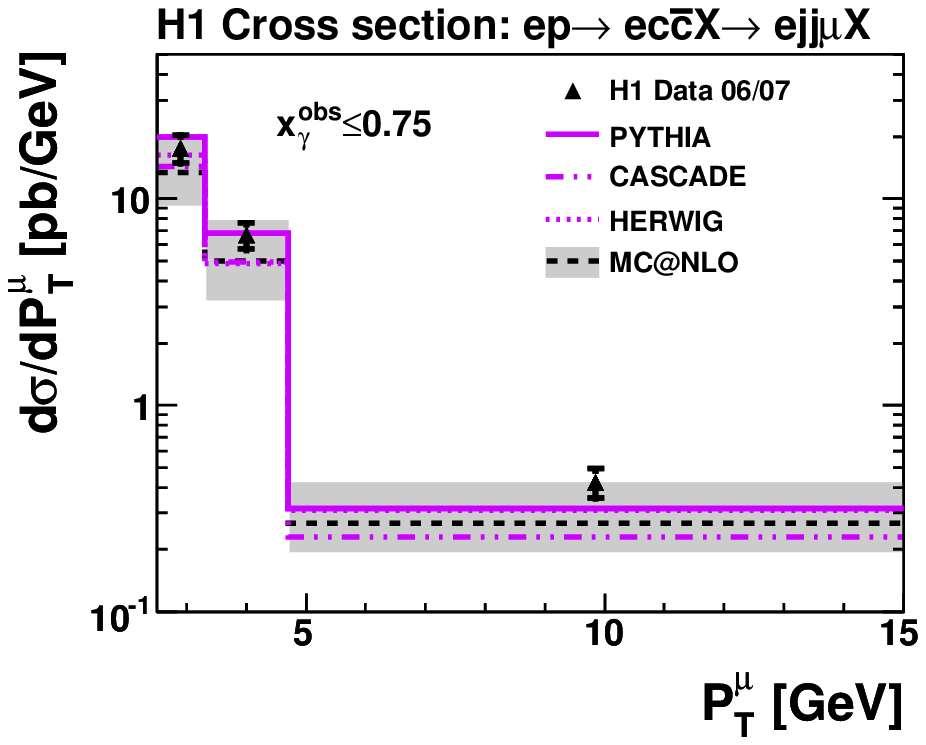,width = 8.5cm}}
\put(7.7,7.5){\epsfig{file=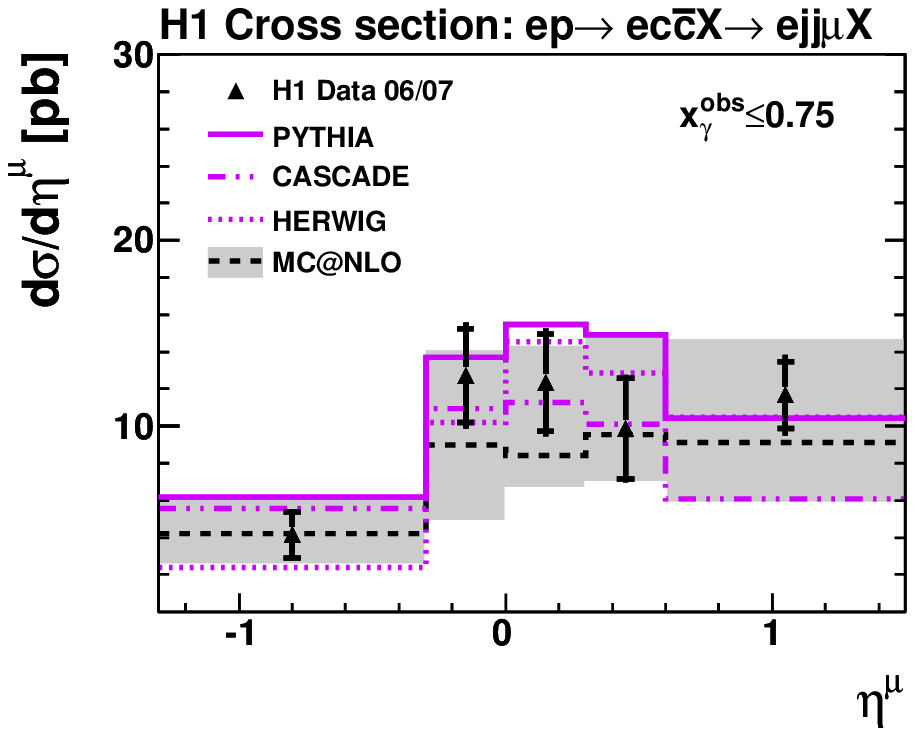,width=8.5cm}}
\put(-1.5,0.){\epsfig{file=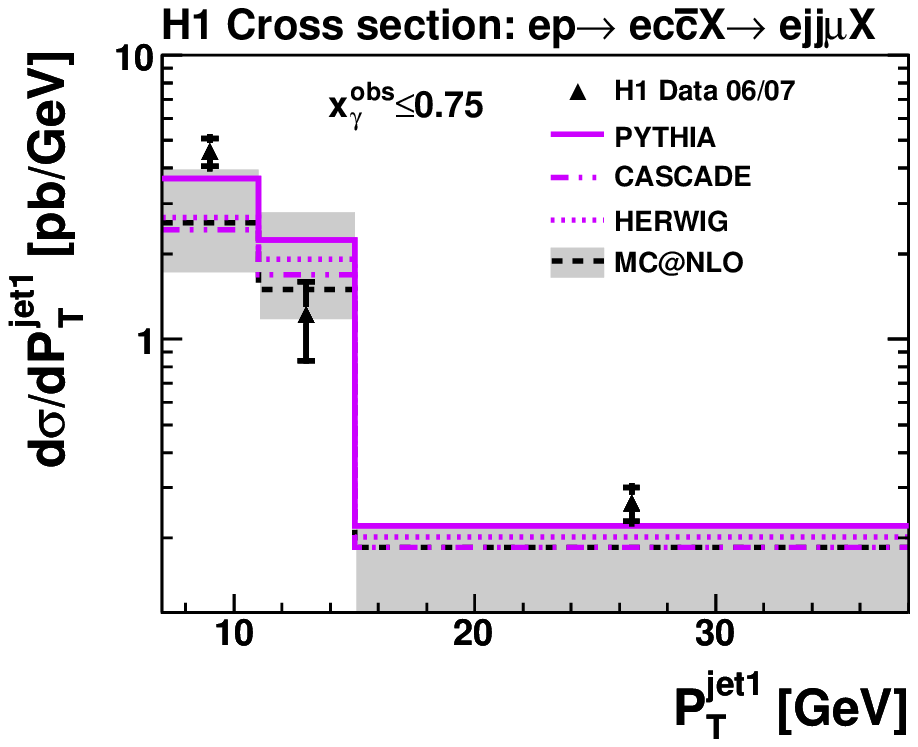,width = 8.5cm}}
\put(7.7,0.){\epsfig{file=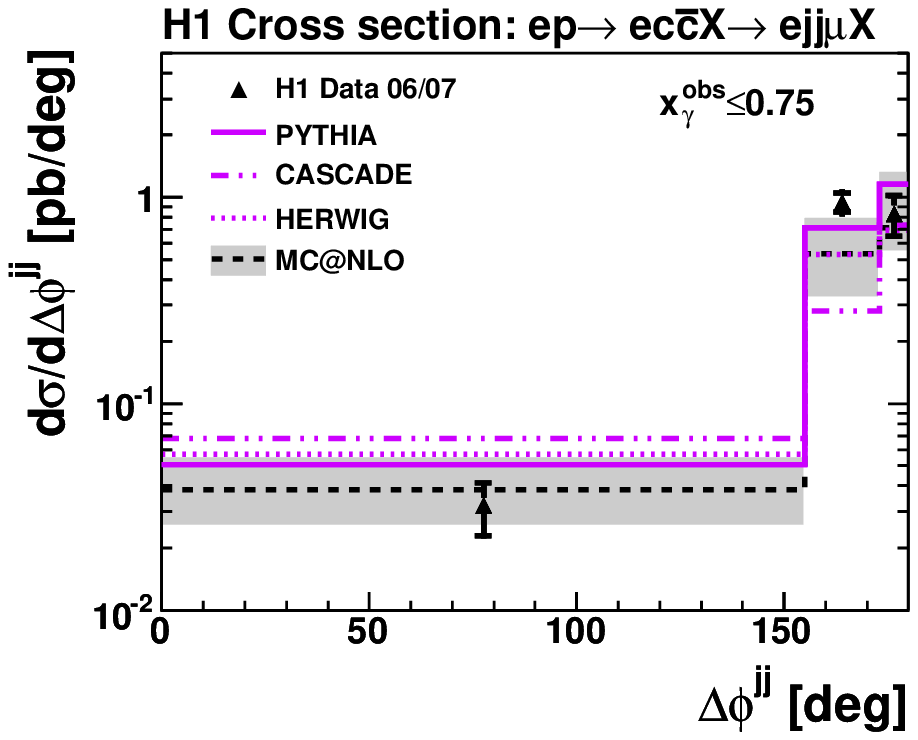,width = 8.5cm}}
\put(0.,14.6){(a)}
\put(9.2,14.6){(b)}
\put(0.,7.2){(c)}
\put(9.2,7.2){(d)}
\end{picture}
\caption{The differential cross sections for charm photoproduction of dijet events using semi-muonic decays for $\xgamma\leq0.75$ as a function of \ptmuon, \etamuon, \ptjetone, and \dphijets.  For details see caption of figure~\ref{fig:beautyxsecfullsample}.}
\label{fig:charmxsecresolvedsample}
\end{figure}

The measured charm cross sections are presented in tables \ref{tab:charmfull}-\ref{tab:charmresolved} and figures \ref{fig:charmxsecfullsample}-\ref{fig:charmxsecresolvedsample}. The distributions are reasonably well described by all models.  
Similar to recent observations in H1 measurements of the photoproduction of $D^*$ mesons \cite{2011DStar}, the central value of the MC@NLO calculations tend to be lower than the measured charm cross sections.

%%%%%%%%%%%%%%%%%%%%%%%%%%%%%%%%%%%%%%%%%%%%%%%%%%%%%%%%%%
\section{Conclusions}
%%%%%%%%%%%%%%%%%%%%%%%%%%%%%%%%%%%%%%%%%%%%%%%%%%%%%%%%%%
Beauty and charm photoproduction cross sections for events with dijets and a muon are measured using the data collected by the H1 detector at HERA. Compared to the previous H1 beauty measurement \cite{H1Hera1desy05004}, the analysis profits from a three times larger luminosity of the data sample, an extended phase space as well as improved understanding of the H1 vertex detector.  The flavour composition of the event sample is determined by the transverse momentum of the muon relative to the jet axis of its associated jet \ptrel\ and by its impact parameter \impactpar. Total visible  and differential cross sections are measured, and the results are compared to leading order QCD models provided by {\sc Pythia}, {\sc Cascade}, and {\sc Herwig}, as well as to the next-to-leading order  calculations provided by MC@NLO. At low values of \ptmuon\ and \ptjetone, the present beauty measurement does not show a significant excess as observed by the previous H1 measurement with respect to the NLO calculation. In general the predictions are in reasonable agreement with the beauty and charm measurements.

%%%%%%%%%%%%%%%%%%%%%%%%%%%%%%%%%%%%%%%%%%%%%%%%%%%%%%%%%%%%
\section*{Acknowledgements}
%%%%%%%%%%%%%%%%%%%%%%%%%%%%%%%%%%%%%%%%%%%%%%%%%%%%%%%%%%

We are grateful to the HERA machine group whose outstanding
efforts have made this experiment possible. 
We thank the engineers and technicians for their work in constructing and
maintaining the H1 detector, our funding agencies for 
financial support, the
DESY technical staff for continual assistance
and the DESY directorate for support and for the
hospitality which they extend to the non-DESY 
members of the collaboration. We would like to thank T. Toll for fruitful discussions and for his help with the MC@NLO predictions.
%%%%%%%%%%%%%%%%%%%%%%%%%%%%%%%%%%%%%%%%%%%%%%%%%%%%%%%%%%%%

\clearpage

%%%%%%%%%%%%%%%%%%%%%%%%%%%%%%%%%%%%%%%%%%%%%%

\renewcommand{\arraystretch}{1.19} % this increase the table in vertical direction, very useful!
\begin{table}[tb]
  \begin{center}
    \begin{tabular}{|rr|rrr||c|}
      \hline
      \multicolumn{6}{|c|}{\bf \boldmath H1 Beauty Dijet Muon Cross Sections}   \\
      \multicolumn{6}{|c|}{$\sigma_{vis}(ep\rightarrow eb\bar{b}X\rightarrow ejj\mu X)$}   \\
      \hline
      \hline
      \multicolumn{2}{|c|}{\ptmuon\ range} & \dsdx{\ptmuon} & stat. & sys. & $f_{b}\pm$stat.  \\  
      \multicolumn{2}{|c|}{[G\eV]} & \multicolumn{3}{c||}{[pb/G\eV]} &  \\ 
      \hline
      $2.5$ & $3.3$ & $25.3$ & $2.2$ & $2.8$ & $0.21\pm0.02$ \\ 
     $3.3$ & $4.7$ & $12.0$ & $0.9$ & $1.3$ & $0.28\pm0.02$\\ 
     $4.7$ & $15.0$ & $0.772$ & $0.078$ & $0.085$ & $0.31\pm0.03$ \\
      \hline
      \hline
      \multicolumn{2}{|c|}{\etamuon\ range} & \dsdx{\etamuon} & stat. & sys. & $f_{b}\pm$stat.   \\  
      \multicolumn{2}{|c|}{} & \multicolumn{3}{c||}{[pb]} &  \\ 
      \hline
      $-1.3$ & $-0.3$ & $9.9$ & $1.2$ & $1.1$ & $0.23\pm0.03$ \\ 
     $-0.3$ & $0.0$ & $19.6$ & $2.2$ & $2.2$ & $0.24\pm0.03$  \\ 
     $0.0$ & $0.3$ & $21.7$ & $2.3$ & $2.4$ & $0.25\pm0.03$ \\ 
     $0.3$ & $0.6$ & $23.4$ & $2.4$ &$2.6$ & $0.30\pm0.03$ \\ 
     $0.6$ & $1.5$ & $14.3$ & $1.4$ & $1.6$ & $0.26\pm0.03$ \\ 
      \hline
      \hline
      \multicolumn{2}{|c|}{\ptjetone\ range} & \dsdx{\ptjetone} & stat. & sys. & $f_{b}\pm$stat.  \\  
      \multicolumn{2}{|c|}{[G\eV]} & \multicolumn{3}{c||}{[pb/G\eV]} &  \\ 
      \hline
     $7$ & $11$ & $4.53$ & $0.41$ & $0.50$ & $0.24\pm0.02$  \\ 
    $11$ & $15$& $3.40$ & $0.27$ & $0.37$ & $0.25\pm0.02$ \\ 
    $15$ & $38$ & $0.469$ & $0.038$ & $0.052$ & $0.30\pm0.02$  \\
      \hline
      \hline
      \multicolumn{2}{|c|}{\dphijets\ range} & \dsdx{\dphijets} & stat. & sys. & $f_{b}\pm$stat.  \\  
      \multicolumn{2}{|c|}{[deg]} & \multicolumn{3}{c||}{[pb/deg]} &  \\ 
      \hline
   $0$ & $155$ & $0.0576$ & $0.0063$ & $0.0061$ & $0.23\pm0.03$ \\ 
$155$ & $173$ & $1.01$ & $0.07$ & $0.11$ & $0.26\pm0.02$  \\ 
$173$ & $180$ & $2.17$ & $0.17$ & $0.24$ & $0.26\pm0.02$ \\ 
      \hline
      \hline
      \multicolumn{2}{|c|}{\xgamma\ range} & \dsdx{\xgamma} & stat. & sys. & $f_{b}\pm$stat.  \\  
      \multicolumn{2}{|c|}{} & \multicolumn{3}{c||}{[pb]} &  \\ 
      \hline
$0.0$ & $0.4$ & $10.6$ & $2.7$ & $1.2$ & $0.17\pm0.04$  \\ 
$0.4 $ & $ 0.75 $ & $ 35.6 $ & $ 3.4 $ & $ 3.9 $ & $ 0.23\pm0.02$   \\ 
$0.75 $ & $ 1.0 $ & $ 103.5 $ & $ 5.9 $ & $ 11.4 $ & $ 0.29\pm0.02$ \\ 
      \hline
    \end{tabular}
    \caption{Bin averaged differential cross sections for beauty photoproduction of dijet events using semi-muonic decays in bins of \ptmuon, \etamuon, \ptjetone, \dphijets, and \xgamma\  with their statistical and systematic uncertainties. The fit parameter $f_b$ is given including its statistical error. }
    \label{tab:beautyfull}
  \end{center}
\end{table}

\begin{table}[tb]
  \begin{center}
    \begin{tabular}{|rr|rrr||c|}
      \hline
      \multicolumn{6}{|c|}{\bf \boldmath H1 Beauty Dijet Muon Cross Sections, \bf \boldmath \xgamma$>0.75$}   \\
      \multicolumn{6}{|c|}{$\sigma_{vis}(ep\rightarrow eb\bar{b}X\rightarrow ejj\mu X)$}   \\
      \hline
      \hline
      \multicolumn{2}{|c|}{\ptmuon\ range} & \dsdx{\ptmuon} & stat. & sys. & $f_{b}\pm$stat.  \\  
      \multicolumn{2}{|c|}{[G\eV]} & \multicolumn{3}{c||}{[pb/G\eV]} &  \\ 
      \hline
     $ 2.5 $ & $ 3.3 $ & $ 14.8 $ & $ 1.5 $ & $ 1.6 $ & $ 0.24\pm0.02 $  \\ 
$3.3 $ & $ 4.7 $ & $ 6.7 $ & $ 0.6 $ & $ 0.7 $ & $ 0.28\pm0.03 $  \\ 
$4.7 $ & $ 15.0 $ & $ 0.487 $ & $ 0.068 $ & $ 0.054 $ & $ 0.31\pm0.04 $ \\
      \hline
      \hline
      \multicolumn{2}{|c|}{\etamuon\ range} & \dsdx{\etamuon} & stat. & sys. & $f_{b}\pm$stat.  \\  
      \multicolumn{2}{|c|}{} & \multicolumn{3}{c||}{[pb]} &  \\ 
      \hline
      $-1.3 $ & $ -0.3 $ & $ 6.5 $ & $ 0.9 $ & $ 0.7 $ & $0.24\pm0.03$  \\ 
$-0.3 $ & $ 0.0 $ & $ 11.3 $ & $ 1.6 $ & $ 1.2 $ & $0.24\pm0.03$ \\ 
$0.0 $ & $ 0.3 $ & $ 14.4 $ & $ 1.8 $ & $ 1.6 $ & $0.29\pm0.03$ \\ 
$0.3 $ & $ 0.6 $ & $ 13.9 $ & $ 1.7 $ & $ 1.5 $ & $0.32\pm0.04$  \\ 
$0.6 $ & $ 1.5 $ & $ 6.1 $ & $ 1.0 $ & $ 0.7 $ & $0.25\pm0.04$ \\ 
      \hline
      \hline
      \multicolumn{2}{|c|}{\ptjetone\ range} & \dsdx{\ptjetone} & stat. & sys. & $f_{b}\pm$stat.   \\  
      \multicolumn{2}{|c|}{[G\eV]} & \multicolumn{3}{c||}{[pb/G\eV]} &  \\ 
      \hline
     $ 7 $ & $ 11 $ & $ 1.81 $ & $ 0.24 $ & $ 0.20 $ & $0.21\pm0.03$ \\ 
$11 $ & $ 15 $ & $ 2.05 $ & $ 0.21 $ & $ 0.23 $ & $0.26\pm0.03$ \\ 
$15 $ & $ 38 $ & $ 0.361 $ & $ 0.032 $ & $ 0.040 $ & $0.35\pm0.03$ \\ 
      \hline
      \hline
      \multicolumn{2}{|c|}{\dphijets\ range} & \dsdx{\dphijets} & stat. & sys. & $f_{b}\pm$stat.  \\  
      \multicolumn{2}{|c|}{[deg]} & \multicolumn{3}{c||}{[pb/deg]} & \\ 
      \hline
 $  0 $ & $ 155 $ & $ 0.0084 $ & $ 0.0030 $ & $ 0.0016 $ & $0.18\pm0.04$ \\ 
$155 $ & $ 173 $ & $ 0.617 $ & $ 0.052 $ & $ 0.068 $ & $0.28\pm0.02$  \\ 
$173 $ & $ 180 $ & $ 1.640 $ & $ 0.144 $ & $ 0.180 $ & $0.30\pm0.03$  \\ 
      \hline
    \end{tabular}
    \caption{Bin averaged differential cross sections for beauty photoproduction of dijet events using semi-muonic decays for $\xgamma>0.75$ in bins of \ptmuon, \etamuon, \ptjetone, and \dphijets \ with their statistical and systematic uncertainties. The fit parameter $f_b$ is given including its statistical error. }
    \label{tab:beautydirect}
  \end{center}
\end{table}

\begin{table}[tb]
  \begin{center}
    \begin{tabular}{|rr|rrr||c|}
      \hline
     \multicolumn{6}{|c|}{\bf \boldmath H1 Beauty Dijet Muon Cross Sections, \bf \boldmath \xgamma$\leq0.75$}   \\
      \multicolumn{6}{|c|}{$\sigma_{vis}(ep\rightarrow eb\bar{b}X\rightarrow ejj\mu X)$}   \\
      \hline
      \hline
      \multicolumn{2}{|c|}{\ptmuon\ range} & \dsdx{\ptmuon} & stat. & sys. & $f_{b}\pm$stat. \\  
      \multicolumn{2}{|c|}{[G\eV]} & \multicolumn{3}{c||}{[pb/G\eV]} &  \\ 
      \hline
      $2.5 $ & $ 3.3 $ & $ 11.3 $ & $ 1.7 $ & $ 1.2 $ & $0.18\pm0.03$ \\ 
$3.3 $ & $ 4.7 $ & $ 5.47 $ & $ 0.67 $ & $ 0.60 $ & $0.29\pm0.03$ \\ 
$4.7 $ & $ 15.0 $ & $ 0.243 $ & $ 0.053 $ & $ 0.027 $ & $0.27\pm0.06$ \\
      \hline
      \hline
      \multicolumn{2}{|c|}{\etamuon\ range} & \dsdx{\etamuon} & stat. & sys. & $f_{b}\pm$stat.   \\  
      \multicolumn{2}{|c|}{} & \multicolumn{3}{c||}{[pb]} &  \\ 
      \hline
      $-1.3 $ & $ -0.3 $ & $ 2.58 $ & $ 1.38 $ & $ 0.29 $ & $ 0.15\pm0.08$ \\ 
$-0.3 $ & $ 0.0 $ & $ 7.98 $ & $ 1.55 $ & $ 0.88 $ & $ 0.23\pm0.04$  \\ 
$0.0 $ & $ 0.3 $ & $ 5.51 $ & $ 1.64 $ & $ 0.61 $ & $ 0.14\pm0.04$ \\ 
$0.3 $ & $ 0.6 $ & $ 7.68 $ & $ 1.77 $ & $ 0.85 $ & $ 0.22\pm0.05$  \\ 
$0.6 $ & $ 1.5 $ & $ 6.38 $ & $ 1.08 $ & $ 0.70 $ & $ 0.21\pm0.04$ \\
      \hline
      \hline
      \multicolumn{2}{|c|}{\ptjetone\ range} & \dsdx{\ptjetone} & stat. & sys. & $f_{b}\pm$stat.  \\  
      \multicolumn{2}{|c|}{[G\eV]} & \multicolumn{3}{c||}{[pb/G\eV]} &  \\ 
      \hline
      $7 $ & $ 11 $ & $ 2.82 $ & $ 0.33 $ & $ 0.31 $ & $0.26\pm0.03$ \\ 
$11 $ & $ 15 $ & $ 1.116 $ & $ 0.188 $ & $ 0.123 $ & $0.20\pm0.03$ \\ 
$15 $ & $ 38 $ & $ 0.095 $ & $ 0.018 $ & $ 0.010 $ & $0.19\pm0.04$  \\
      \hline
      \hline
      \multicolumn{2}{|c|}{\dphijets\ range} & \dsdx{\dphijets} & stat. & sys. & $f_{b}\pm$stat.   \\  
      \multicolumn{2}{|c|}{[deg]} & \multicolumn{3}{c||}{[pb/deg]} &  \\ 
      \hline
     $ 0 $ & $ 155 $ & $ 0.0365 $ & $ 0.0063 $ & $ 0.0040 $ & $0.23\pm0.04$ \\ 
$155 $ & $ 173 $ & $ 0.462 $ & $ 0.056 $ & $ 0.051 $ & $0.26\pm0.03$ \\ 
$173 $ & $ 180 $ & $ 0.320 $ & $ 0.012 $ & $ 0.035 $ & $0.11\pm0.04$  \\
      \hline
    \end{tabular}
    \caption{Bin averaged differential cross sections for beauty photoproduction of dijet events using semi-muonic decays for $\xgamma\leq0.75$ in bins of \ptmuon, \etamuon, \ptjetone, and \dphijets \ with their statistical and systematic uncertainties. The fit parameter $f_b$ is given including its statistical error. }
    \label{tab:beautyresolved}
  \end{center}
\end{table}

\begin{table}[tb]
  \begin{center}
    \begin{tabular}{|rr|rrr||c|}
      \hline
      \multicolumn{6}{|c|}{\bf \boldmath H1 Charm Dijet Muon Cross Sections}   \\
      \multicolumn{6}{|c|}{$\sigma_{vis}(ep\rightarrow ec\bar{c}X\rightarrow ejj\mu X)$}   \\
      \hline
      \hline
      \multicolumn{2}{|c|}{\ptmuon\ range} & \dsdx{\ptmuon} & stat. & sys. & $f_{c}\pm$stat.  \\  
      \multicolumn{2}{|c|}{[G\eV]} & \multicolumn{3}{c||}{[pb/G\eV]} & \\ 
      \hline
      $2.5 $ & $ 3.3 $ & $ 49.1 $ & $ 4.0 $ & $ 5.1 $ & $0.50\pm0.04$\\ 
$3.3 $ & $ 4.7 $ & $ 18.3 $ & $ 1.7 $ & $ 1.9 $ & $0.47\pm0.04$ \\ 
$4.7 $ & $ 15.0 $ & $ 0.854 $ & $ 0.126 $ & $ 0.088 $ & $0.31\pm0.04$ \\
      \hline
      \hline
      \multicolumn{2}{|c|}{\etamuon\ range} & \dsdx{\etamuon} & stat. & sys. & $f_{c}\pm$stat.  \\  
      \multicolumn{2}{|c|}{} & \multicolumn{3}{c||}{[pb]} &  \\ 
      \hline
      $-1.3 $ & $ -0.3 $ & $ 20.3 $ & $ 2.5 $ & $ 2.1 $ & $0.44\pm0.05$\\ 
$-0.3 $ & $ 0.0 $ & $ 37.0 $ & $ 4.6 $ & $ 3.9 $ & $0.45\pm0.05$ \\ 
$0.0 $ & $ 0.3 $ & $ 42.3 $ & $ 4.6 $ & $ 4.4 $ & $0.48\pm0.05$ \\ 
$0.3 $ & $ 0.6 $ & $ 38.0 $ & $ 4.6 $ & $ 4.0 $ & $0.49\pm0.06$ \\ 
$0.6 $ & $ 1.5 $ & $ 23.6 $ & $ 3.1 $ & $ 2.5 $ & $0.46\pm0.05$ \\
      \hline
      \hline
      \multicolumn{2}{|c|}{\ptjetone\ range} & \dsdx{\ptjetone} & stat. & sys. & $f_{c}\pm$stat.  \\  
      \multicolumn{2}{|c|}{[G\eV]} & \multicolumn{3}{c||}{[pb/G\eV]} & \\ 
      \hline
      $7 $ & $ 11 $ & $ 11.8 $ & $ 0.9 $ & $ 1.2 $ & $0.55\pm0.04$\\ 
$11 $ & $ 15 $ & $ 5.22 $ & $ 0.59 $ & $ 0.54 $ & $0.39\pm0.04$ \\ 
$15 $ & $ 38 $ & $ 0.657 $ & $ 0.066 $ & $ 0.068 $ & $0.51\pm0.05$ \\ 

      \hline
      \hline
      \multicolumn{2}{|c|}{\dphijets\ range} & \dsdx{\dphijets} & stat. & sys. & $f_{c}\pm$stat.   \\  
      \multicolumn{2}{|c|}{[deg]} & \multicolumn{3}{c||}{[pb/deg]} & \\ 
      \hline
     $0 $ & $ 155 $ & $ 0.0820 $ & $ 0.0119 $ & $ 0.0085 $ & $0.37\pm0.05$ \\ 
$155 $ & $ 173 $ & $ 2.10 $ & $ 0.15 $ & $ 0.22 $ & $0.52\pm0.04$ \\ 
$173 $ & $ 180 $ & $ 4.06 $ & $ 0.35 $ & $ 0.43 $ & $0.50\pm0.04$ \\ 

      \hline
      \hline
      \multicolumn{2}{|c|}{\xgamma\ range} & \dsdx{\xgamma} & stat. & sys. & $f_{c}\pm$stat. \\  
      \multicolumn{2}{|c|}{} & \multicolumn{3}{c||}{[pb]} & \\ 
      \hline
     $ 0.0 $ & $ 0.4 $ & $ 12.3 $ & $ 5.0 $ & $ 1.3 $ & $0.22\pm0.09$ \\ 
$0.4 $ & $ 0.75 $ & $ 63.5 $ & $ 5.8 $ & $ 6.6 $ & $0.50\pm0.04$ \\ 
$0.75 $ & $ 1.0 $ & $ 206.7 $ & $ 13.8 $ & $ 21.5 $ & $0.51\pm0.03$ \\
      \hline
    \end{tabular}
    \caption{Bin averaged differential cross sections for charm photoproduction of dijet events using semi-muonic decays in bins of \ptmuon, \etamuon, \ptjetone, \dphijets, and \xgamma \ with their statistical and systematic uncertainties. The fit parameter $f_c$ is given including its statistical error. }
    \label{tab:charmfull}
  \end{center}
\end{table}

\begin{table}[tb]
  \begin{center}
    \begin{tabular}{|rr|rrr||c|}
      \hline
      \multicolumn{6}{|c|}{\bf \boldmath H1 Charm Dijet Muon Cross Sections, \bf \boldmath \xgamma$>0.75$}   \\
      \multicolumn{6}{|c|}{$\sigma_{vis}(ep\rightarrow ec\bar{c}X\rightarrow ejj\mu X)$}   \\
      \hline
      \hline
      \multicolumn{2}{|c|}{\ptmuon\ range} & \dsdx{\ptmuon} & stat. & sys. & $f_{c}\pm$stat.  \\  
      \multicolumn{2}{|c|}{[G\eV]} & \multicolumn{3}{c||}{[pb/G\eV]} & \\ 
      \hline
     $2.5 $ & $ 3.3 $ & $ 31.2 $ & $ 3.2 $ & $ 3.2 $ & $0.55\pm0.05$  \\ 
$3.3 $ & $ 4.7 $ & $ 11.6 $ & $ 1.4 $ & $ 1.2 $ & $0.49\pm0.06$ \\ 
$4.7 $ & $ 15.0 $ & $ 0.368 $ & $ 0.119 $ & $ 0.038 $ & $0.20\pm0.07$ \\ 

      \hline
      \hline
      \multicolumn{2}{|c|}{\etamuon\ range} & \dsdx{\etamuon} & stat. & sys. & $f_{c}\pm$stat.  \\  
      \multicolumn{2}{|c|}{} & \multicolumn{3}{c||}{[pb]} & \\ 
      \hline
     $ -1.3 $ & $ -0.3 $ & $ 14.6 $ & $ 2.2 $ & $ 1.5 $ & $0.45\pm0.06$ \\ 
$-0.3 $ & $ 0.0 $ & $ 21.9 $ & $ 4.0 $ & $ 2.3 $ & $0.41\pm0.07$ \\ 
$0.0 $ & $ 0.3 $ & $ 27.0 $ & $ 3.4 $ & $ 2.8 $ & $0.50\pm0.06$ \\ 
$0.3 $ & $ 0.6 $ & $ 25.8 $ & $ 3.0 $ & $ 2.7 $ & $0.56\pm0.06$ \\ 
$0.6 $ & $ 1.5 $ & $ 9.4 $ & $ 1.9 $ & $1.0 $ & $0.39\pm0.08$\\ 
      \hline
      \hline
      \multicolumn{2}{|c|}{\ptjetone\ range} & \dsdx{\ptjetone} & stat. & sys. & $f_{c}\pm$stat.  \\  
      \multicolumn{2}{|c|}{[G\eV]} & \multicolumn{3}{c||}{[pb/G\eV]} & \\ 
      \hline
     $ 7 $ & $ 11 $ & $ 6.6 $ & $ 0.6 $ & $ 0.7 $ & $0.57\pm0.05$ \\ 
$11 $ & $ 15 $ & $ 3.45 $ & $ 0.42 $ & $ 0.36 $ & $0.42\pm0.05$ \\ 
$15 $ & $ 38 $ & $ 0.377 $ & $ 0.048 $ & $ 0.039 $ & $0.42\pm0.05$ \\ 
      \hline
      \hline
      \multicolumn{2}{|c|}{\dphijets\ range} & \dsdx{\dphijets} & stat. & sys. & $f_{c}\pm$stat.  \\  
      \multicolumn{2}{|c|}{[deg]} & \multicolumn{3}{c||}{[pb/deg]} & \\ 
      \hline
     $0 $ & $ 155 $ & $ 0.0424 $ & $ 0.0075 $ & $ 0.0045 $ & $0.46\pm0.08$ \\ 
$155 $ & $ 173 $ & $ 1.202 $ & $ 0.122 $ & $ 0.125 $ & $0.46\pm0.05$ \\ 
$173 $ & $ 180 $ & $ 3.00 $ & $ 0.29 $ & $ 0.31 $ & $0.53\pm0.05$ \\ 

      \hline
    \end{tabular}
    \caption{Bin averaged differential cross sections for charm photoproduction of dijet events using semi-muonic decays for $\xgamma>0.75$ in bins of \ptmuon, \etamuon, \ptjetone, and \dphijets \ with their statistical and systematic uncertainties. The fit parameter $f_c$ is given including its statistical error. }
    \label{tab:charmdirect}
  \end{center}
\end{table}

\begin{table}[tb]
  \begin{center}
    \begin{tabular}{|rr|rrr||c|}
      \hline
       \multicolumn{6}{|c|}{\bf \boldmath H1 Charm Dijet Muon Cross Sections, \bf \boldmath \xgamma$\leq0.75$}   \\
      \multicolumn{6}{|c|}{$\sigma_{vis}(ep\rightarrow ec\bar{c}X\rightarrow ejj\mu X)$}   \\
      \hline
      \hline
      \multicolumn{2}{|c|}{\ptmuon\ range} & \dsdx{\ptmuon} & stat. & sys. & $f_{c}\pm$stat.  \\  
      \multicolumn{2}{|c|}{[G\eV]} & \multicolumn{3}{c||}{[pb/G\eV]} & \\ 
      \hline
      $2.5 $ & $ 3.3 $ & $ 17.6 $ & $ 2.7 $ & $ 1.8 $ & $0.43\pm0.06$ \\ 
$3.3 $ & $ 4.7 $ & $ 6.7 $ & $ 1.0 $ & $ 0.7 $ & $0.45\pm0.06$\\ 
$4.7 $ & $ 15.0 $ & $ 0.424 $ & $ 0.068 $ & $ 0.044 $ & $0.45\pm0.07$ \\
      \hline
      \hline
      \multicolumn{2}{|c|}{\etamuon\ range} & \dsdx{\etamuon} & stat. & sys. & $f_{c}\pm$stat.  \\  
      \multicolumn{2}{|c|}{} & \multicolumn{3}{c||}{[pb]} & \\ 
      \hline
     $ -1.3 $ & $ -0.3 $ & $ 4.16 $ & $ 1.25 $ & $ 0.43 $ & $0.32\pm0.10$\\ 
$-0.3 $ & $ 0.0 $ & $ 12.7 $ & $ 2.5 $ & $ 1.3 $ & $0.45\pm0.08$ \\ 
$0.0 $ & $ 0.3 $ & $ 12.3 $ & $ 2.6 $ & $ 1.3 $ & $0.37\pm0.08$ \\
$0.3 $ & $ 0.6 $ & $ 9.9 $ & $ 2.7 $ & $ 1.0 $ & $0.32\pm0.09$ \\ 
$0.6 $ & $ 1.5 $ & $ 11.7 $ & $ 1.8 $ & $ 1.2 $ & $0.43\pm0.06$ \\
      \hline
      \hline
      \multicolumn{2}{|c|}{\ptjetone\ range} & \dsdx{\ptjetone} & stat. & sys. & $f_{c}\pm$stat.  \\  
      \multicolumn{2}{|c|}{[G\eV]} & \multicolumn{3}{c||}{[pb/G\eV]} &\\ 
      \hline
      $7 $ & $ 11 $ & $ 4.58 $ & $ 0.51 $ & $ 0.48 $ & $0.50\pm0.05$ \\ 
$11 $ & $ 15 $ & $ 1.216 $ & $ 0.375 $ & $ 0.126 $ & $0.24\pm0.07$ \\ 
$15 $ & $ 38 $ & $ 0.265 $ & $ 0.037 $ & $ 0.028 $ & $0.65\pm0.08$  \\
      \hline
      \hline
      \multicolumn{2}{|c|}{\dphijets\ range} & \dsdx{\dphijets} & stat. & sys. & $f_{c}\pm$stat.  \\  
      \multicolumn{2}{|c|}{[deg]} & \multicolumn{3}{c||}{[pb/deg]} & \\ 
      \hline
      $0 $ & $ 155 $ & $ 0.0320 $ & $ 0.0092 $ & $ 0.0047 $ & $0.24\pm0.07$ \\ 
$155 $ & $ 173 $ & $ 0.947 $ & $ 0.102 $ & $ 0.098 $ & $0.64\pm0.07$\\ 
$173 $ & $ 180 $ & $ 0.834 $ & $ 0.185 $ & $ 0.087 $ & $0.35\pm0.07$ \\ 

      \hline
    \end{tabular}
    \caption{Bin averaged differential cross sections for charm photoproduction of dijet events using semi-muonic decays for $\xgamma\leq0.75$ in bins of \ptmuon, \etamuon, \ptjetone, and \dphijets \ with their statistical and systematic uncertainties. The fit parameter $f_c$ is given including its statistical error. }
    \label{tab:charmresolved}
  \end{center}
\end{table}

%%%%%%%%%%%%%%%%%%%%%%%%%%%%%%%%%%%%%%%%%%%%%%%%%%%%%%%%%%%

\clearpage

\clearpage

\clearpage

\clearpage

\clearpage

\end{document}